\documentclass[twocolumn]{aastex61}
\usepackage{color}

\newcommand\aastex{AAS\TeX}

\def\mdot{\hbox{M$_{\odot}$}} 
\def\pers{\hbox{s$^{-1}$}}

\shorttitle{\aastex\ Spectroscopic Orbits of Eleven M Dwarf Binaries}
\shortauthors{Winters et al.}

\begin{document}

\title{Spectroscopic Orbits of Eleven Nearby, Mid-to-Late M Dwarf Binaries}

\correspondingauthor{Jennifer G. Winters}
\email{jennifer.winters@cfa.harvard.edu}

\author[0000-0001-6031-9513]{Jennifer G. Winters}
\affil{Harvard-Smithsonian Center for Astrophysics, 60 Garden Street,
  Cambridge, MA 02138, USA}

\author{Jonathan M. Irwin}
\affil{Harvard-Smithsonian Center for Astrophysics, 60 Garden Street,
  Cambridge, MA 02138, USA}

\author[0000-0002-9003-484X]{David Charbonneau}
\affil{Harvard-Smithsonian Center for Astrophysics, 60 Garden Street,
  Cambridge, MA 02138, USA}

\author[0000-0001-9911-7388]{David W. Latham}
\affil{Harvard-Smithsonian Center for Astrophysics, 60 Garden Street,
  Cambridge, MA 02138, USA}
  
\author[0000-0001-8726-3134]{Amber M. Medina}
\affil{Harvard-Smithsonian Center for Astrophysics, 60 Garden Street,
  Cambridge, MA 02138, USA}

\author{Jessica Mink}
\affil{Harvard-Smithsonian Center for Astrophysics, 60 Garden Street,
  Cambridge, MA 02138, USA}

\author[0000-0002-9789-5474]{Gilbert A. Esquerdo}
\affil{Harvard-Smithsonian Center for Astrophysics, 60 Garden Street,
  Cambridge, MA 02138, USA}

\author{Perry Berlind}
\affil{Harvard-Smithsonian Center for Astrophysics, 60 Garden Street,
  Cambridge, MA 02138, USA}

\author{Michael L. Calkins}
\affil{Harvard-Smithsonian Center for Astrophysics, 60 Garden Street,
  Cambridge, MA 02138, USA}

\author[0000-0002-3321-4924]{Zachory K. Berta-Thompson}
\affil{Department of Astrophysical and Planetary Sciences, University of Colorado, Boulder, CO 80309, USA}

\begin{abstract}

We present the spectroscopic orbits of eleven nearby, mid-to-late M
dwarf binary systems in a variety of configurations: two single-lined
binaries (SB1s), seven double-lined binaries (SB2s), one double-lined
triple (ST2), and one triple-lined triple (ST3). Eight of these orbits
are the first published for these systems, while five are newly
identified multiples. We obtained multi-epoch, high-resolution spectra
with the TRES instrument on the 1.5m Tillinghast Reflector at the Fred
Lawrence Whipple Observatory located on Mt. Hopkins in AZ. Using the
TiO molecular bands at $7065 -- 7165$~\AA, we calculated radial
velocities for these systems, from which we derived their orbits. We
find LHS~1817 to have in a 7-hour period a companion that is likely a
white dwarf, due to the ellipsoidal modulation we see in our
MEarth-North light curve data. We find G~123-45 and LTT~11586 to host
companions with minimum masses of 41~$M_{Jup}$ and 44~$M_{Jup}$ with
orbital periods of 35 and 15 days, respectively. We find
2MA~0930$+$0227 to have a rapidly rotating stellar companion in a
917-day orbital period. GJ~268, GJ~1029, LP~734-34, GJ~1182, G~258-17,
and LTT~7077 are SB2s with stellar companions with orbital periods of
10, 96, 34, 154, 5, and 84 days; LP~655-43 is an ST3 with one
companion in an 18-day orbital period and an outer component in a
longer undetermined period. In addition, we present radial velocities
for both components of L~870-44AB and for the outer components of
LTT~11586 and LP~655-43.

\end{abstract}

\keywords{stars: low-mass -- binaries: spectroscopic -- binaries (including multiple): close -- stars: kinematics and dynamics}

\section{Introduction} \label{sec:intro}

Stellar multiplicity is known to be a decreasing function of primary
mass, where low-mass main sequence stars have stellar companions less
frequently than more massive stars. The multiplicity rate (MR) is
  the percentage of all systems with a particular primary mass that
  are multiple, regardless of whether the system is double, triple, or
  higher order. Solar-type stars have an MR of roughly 46\%
\citep{Raghavan(2010)}, ~while the MR for M dwarfs appears to be
converging on values around 27\%
\citep{Duchene(2013),Ward-Duong(2015),Winters(2019a)}. Suggestions for
this observed mass dependence include the dynamical stripping of
companions from their more massive primaries on gigayear timescales
and/or a metallicity dependence in the binary formation environment
\citep{Duchene(2013)}; this topic remains an area of active research.

However, the closest M dwarfs have not been surveyed comprehensively
with the high-resolution techniques necessary to detect companions at
the smallest separations, so our understanding of the distribution of
the orbital parameters (periods, mass ratios, separations,
eccentricities) of M dwarf binaries remains incomplete. Orbital
measurements provide important constraints to binary star formation
and evolutionary models at the low-mass end of the stellar main
sequence. For example, binary star formation models can be probed
  by fitting the shape of the orbital period
distribution. Differentiating between a log-normal or a power law
  fit would illuminate whether star formation has a preferred spatial
scale or follows a scale-free process, respectively, in the formation
of binaries \citep{Duchene(2013)}. Stars with large space motions are
typically old, while slow-moving stars are typically
young. Evolutionary models can be constrained by exploring
multiplicity as a function of space motion by using the gamma velocity
of the system, a valuable result of orbit
measurements. \citet{Winters(2019a)} found a decreasing trend of
stellar multiplicity with increasing tangential velocity for M
dwarfs. The calculation of three dimensional $UVW$ space motions from
the combination of gamma velocities with the proper motions of binary
systems permits a robust investigation of this trend. Further, the
resulting minimum masses from orbit measurements, when combined with
inclinations from astrometric orbits (where the components are
unresolved), yield dynamical masses for these low-mass objects. Yet,
within 25 pc, only a few dozen M dwarf systems have measured
spectroscopic orbits, mostly with early-type M dwarf primaries; within
15~pc, only nine mid-to-late M-dwarf (0.1 -- 0.3 M$_{\odot}$)
multiples have published spectroscopic orbits
\citep{Lacy(1977),Delfosse(1999c),Segransan(2000),Nidever(2002),Baroch(2018)}.

Part of the reason for the dearth of spectroscopic orbits with M dwarf
primaries is that their intrinsically low luminosities have made them
historically challenging targets to study. This faintness is
compounded by the multi-epoch observations required to detect radial
velocity variations. Thus, early spectroscopic work on M dwarfs
focused on bright, typically early-type M dwarf targets
\citep{Duquennoy(1988b),Marcy(1989),Tokovinin(1992b),Mazeh(2003)}.

However, the M dwarf spectral sequence spans a magnitude difference of
11.2 in $M_{V}$ (8.8 -- 20.0 mag) and a factor of 8 in mass (0.08 --
0.64 M/M$_{\odot}$). Thus, the orbital parameters of binaries with
more massive, early-type M dwarf primaries are not necessarily
predictive of those with mid-to-late M dwarf primaries. More work
needs to be done to characterize binaries with later-type
primaries. Fortunately, modern spectrographs now allow us to push
toward fainter targets and larger samples, as illustrated by the work
presented in
\citet{Delfosse(1999c),Delfosse(1999d),Nidever(2002),Shkolnik(2010),Davison(2014),Baroch(2018)}.

We are conducting an all-sky, volume-complete, multi-epoch,
high-resolution spectroscopic survey of 412 mid-to-late M dwarfs (0.1
-- 0.3 M$_{\odot}$) within 15 pc for companions. For targets north of
$\delta$ $=-$15\arcdeg, we are using the Tillinghast Reflector Echelle
Spectrograph (TRES) on the 1.5m telescope at the Fred Lawrence Whipple
Observatory (FLWO) on Mt. Hopkins, AZ. For targets south of $\delta
=-$15\arcdeg, we are using the CTIO HIgh ResolutiON (CHIRON)
spectrograph at the Cerro Tololo Inter-American Observatory / Small
and Moderate Aperture Research Telescope System (CTIO / SMARTS) 1.5m
telescope. During the TRES portion of our survey, we have discovered
six new spectroscopic multiple systems to-date. One system,
LHS~1610Ab, was previously presented in \citet{Winters(2018)}; we do
not include it here. Here we present the spectroscopic orbits of
eleven multiples containing 24 components; in addition, we present the
velocities of L~870-44AB, a very long-period binary that shows doubled
lines, for which we have not measured an orbit.  Future papers from
this project will present the full sample of mid-to-late M dwarfs
within 15 pc, along with their radial and rotational velocities,
H$_{\alpha}$ equivalent widths, $UVW$ space motions, and a thorough
analysis of the multiplicity of this volume-complete sample.

\section{Sample Selection} \label{sec:new_mults}

All of the systems presented here were thought to have primary masses
0.1 -- 0.3 M/M$_{\odot}$ and to lie within 15 pc (corresponding to a
parallax $\pi > 66.67$ mas), either via a trigonometric parallax or a
photometric distance estimate. Results from the {\it Gaia} second data
release (DR2) \citep{GaiaDR2(2018),Lindegren(2018)} provided new or
revised parallaxes for all of them. As expected, systems that were
discovered to be nearly equal-luminosity binaries that previously had
only photometric distance estimates now lie beyond 15 pc via
  their trigonometric parallax, as light from the unresolved
secondary made the system overluminous and resulted in an
underestimated photometric distance. This was the case for three
systems presented here: L~870-44, LP~655-43, LP~734-34. In addition,
G~258-17 had a parallax that placed it beyond 25 pc, of which we
became aware only after we discovered it was a new SB2 and began
measuring its orbit. Table \ref{tab:sample} lists the astrometric data
from the {\it Gaia} DR2 for the systems presented here, where the
coordinates were adjusted for proper motion from the epoch of the
2MASS observations to epoch 2000.0 using the DR2 proper motions
\citep{GaiaDR2(2018)} and the 2MASS right ascenscion, declination, and
Julian date \citep{Skrutskie(2006)} for each object. We additionally
list the best-known distance at the time before the availability
of the DR2, which led to the inclusion of each system in our sample,
as well as the DR2 parallax. The initial parallaxes listed for
GJ~1029AB and GJ~268AB represent weighted mean parallaxes.

Masses were estimated using the M$_{K}$ mass-luminosity relation (MLR)
by \citet{Benedict(2016)} under the assumption that the objects were
single stars. Three exceptions to this are GJ~268AB, LTT~11586AcB, and
LTT~7077AB, which were already reported in the literature to be
spectroscopic binaries. We elected to measure the orbit of GJ~268AB as
a check for our method (see \S \ref{subsubsec:gj0268}) with the
knowledge that the primary's mass was within the range of our sample
\citep[i.e., from the orbit presented in][]{Delfosse(1999c)}. We chose
to measure the spectroscopic orbits of LTT~11586AcB and LTT~7077AB,
systems with no published orbits, under the assumption that they
were nearly-equal-magnitude binaries and that the primary masses from
the measured orbits would then fall within the range of targeted
stellar masses in our sample. Table \ref{tab:phot} provides the
optical and infrared photometric data for the sample, where
available. All photometry measured by Weis
\citep[i.e.,][]{Weis(1991b),Weis(1996),Weis(1999)} has been converted
to the Johnson-Kron-Cousins (JKC) system using the relation in
\citet{Bessell(1987)}. In addition, we list each object's initial mass
estimate (from the initial parallaxes in Table \ref{tab:sample}, in
combination with the $K-$band magnitude in Table \ref{tab:phot}), and
the types of spectroscopic multiple for each system. We note that
  the initial mass estimates are incorrect, as they include light from
  the companion star and/or their initial distances were erroneous. We
  list them here to illustrate how these objects came to be intially
  included in our sample. More accurate masses can be derived from the
  results of our orbit determinations under the assumption that there
  are no more unresolved stars in the system. We also indicate
whether the multiple system is a new discovery (`New Mult?') and
whether the orbit presented here is the first orbit measured for the
system (`First Orbit?'). We note that, aside from GJ~1029AB,
L~870-44AB, GJ~268AB, and GJ~1182AB, these are the first spectroscopic
orbits measured for these systems. In the cases of GJ~1029AB and
GJ~1182AB, we only became aware of their previously published orbits
by \citet{Baroch(2018)} after we had completed our orbit
determinations for those systems. L~870-44AB has no published orbit,
and we do not report one here.

\begin{deluxetable*}{lcclccccr}
\tabletypesize{\scriptsize}
\tablecaption{Astrometric Data for Multiple Systems \label{tab:sample}}
\tablecolumns{9}
\tablenum{1}
\tablewidth{0pt}
\tablehead{\colhead{Name}            &
	   \colhead{2MASS ID}        &
	   \colhead{R.A.}            &
           \colhead{decl.}           &
           \colhead{$\mu_{\rm RA}$}    &
	   \colhead{$\mu_{\rm decl}$}   &
	   \colhead{$\pi_{init}$}      &
	   \colhead{Ref}              &
       \colhead{$\pi_{DR2}$}           \\
	   \colhead{   }              &
	   \colhead{   }              &
           \colhead{(hh:mm:ss)}       &
           \colhead{(dd:mm:ss)}       &
	   \colhead{(mas yr$^{-1}$)}   &
           \colhead{(mas yr$^{-1}$)}   &
	   \colhead{(mas)}            & 
	   \colhead{}                 & 
        \colhead{(mas)}               
}

\startdata
GJ~1029AB          & 01053732$+$2829339 & 01:05:37.6 &$+$28:29:34  & 1914.461 & -188.452 & 83.70$\pm$2.19  & 1,3  & 79.84$\pm$0.34  \\ 
L~870-44AB         & 01463681$-$0838578 & 01:46:36.8 &$-$08:38:58  &  412.840 & -182.424 & 68.49$\pm$10.6* & 5    & 39.10$\pm$1.05  \\ 
LP~655-43ABC       & 04380252$-$0556132 & 04:38:02.5 &$-$05:56:13  &  -63.909 & -180.631 & 68.03$\pm$10.9* & 5    & 27.19$\pm$0.30  \\ 
LTT~11586AcB       & 05074924$+$1758584 & 05:07:49.2 &$+$17:58:58  &   32.012 & -261.584 &101.90$\pm$6.00  & 2    & 86.10$\pm$0.31  \\ 
LHS~1817Ab         & 06052936$+$6049231 & 06:05:29.4 &$+$60:49:23  &  289.645 & -788.403 & 71.30$\pm$2.20  & 1    & 61.13$\pm$0.15  \\ 
GJ~268AB           & 07100180$+$3831457 & 07:10:01.8 &$+$38:31:46  & -439.670 & -944.785 &163.41$\pm$1.78  & 3,4  &164.64$\pm$0.13  \\
2MA~0930+0227AB    & 09305084$+$0227202 & 09:30:50.8 &$+$02:27:20  &  -33.530 &   80.977 & 71.30$\pm$7.20  & 2    & 44.50$\pm$0.37  \\
LP~734-34AB        & 12102834$-$1310234 & 12:10:28.4 &$-$13:10:24  &  246.653 & -343.832 & 71.94$\pm$11.7* & 5    & 45.45$\pm$0.08  \\ 
G~123-45Ab         & 12362870$+$3512007 & 12:36:28.7 &$+$35:12:01  & -360.030 & -117.185 & 88.20$\pm$1.70  & 1    & 84.01$\pm$0.21  \\ 
GJ~1182AB          & 14153253$+$0439312 & 14:15:32.6 &$+$04:39:31  & -744.931 & -766.435 & 71.70$\pm$3.40  & 3    & 71.11$\pm$0.39  \\ 
G~258-17AB         & 17411611$+$7226320 & 17:41:16.1 &$+$72:26:32  & -124.133 &  301.274 & 77.60$\pm$5.00  & 1    & 33.53$\pm$0.05  \\ 
LTT~7077AB         & 17462934$-$0842362 & 17:46:29.4 &$-$08:42:37  &  -44.202 & -428.041 & 77.18$\pm$1.71  & 6    & 76.59$\pm$0.08  \\      
\enddata
\tablerefs{(1) \citet{Dittmann(2014)}; (2) \citet{Finch(2018)}; (3)
  \citet{vanAltena(1995)}; (4) \citet{vanLeeuwen(2007)}; (5)
  \citet{Winters(2015)}; (6) \citet{Winters(2017)}.}
\tablecomments{Proper motions $\mu_{\rm RA}$, $\mu_{\rm decl}$ and
  parallaxes $\pi_{DR2}$ are from the {\it Gaia} DR2
  \citep{GaiaDR2(2018),Lindegren(2018)}. An * next to the $\pi_{init}$
  indicates that the listed `parallax' is a photometric distance
  estimate.}
\end{deluxetable*}

\begin{deluxetable*}{lccccccccccccc}
\tabletypesize{\scriptsize}
\tablecaption{Photometric and Spectroscopic Data for Multiple Systems \label{tab:phot}}
\tablecolumns{13}
\tablenum{2}
\tablewidth{0pt}
\tablehead{\colhead{Name}                &
	   \colhead{$G$}                 &
	   \colhead{$V_J$}               &
           \colhead{$R_{KC}$}             &
	   \colhead{$I_{KC}$}             &
           \colhead{Ref}                 &
           \colhead{$J$}                 &
           \colhead{$H$}                 &
           \colhead{$K_s$}               &
           \colhead{mass$_{init}$}        &
	   \colhead{Type}               &
	   \colhead{New}                &
	   \colhead{First}              \\
	   \colhead{   }                 &
           \colhead{(mag)}               &
           \colhead{(mag)}               &
	   \colhead{(mag)}               &
           \colhead{(mag)}               &
           \colhead{   }                 &
           \colhead{(mag)}               &
	   \colhead{(mag)}               &
           \colhead{(mag)}               &
           \colhead{(M/M$_{\odot}$)}       &
	   \colhead{ }                   &
	   \colhead{Mult?}               &
	   \colhead{Orbit?}
}

\startdata
GJ~1029AB       &  12.97 &  14.79J & 13.29J & 11.37J &   4   &  9.486J & 8.881J & 8.550J & 0.16$\pm$0.02   & SB2  & no  & no    \\ 
L~870-44AB      &  11.73 &  12.99J & 11.82J & 10.30J &   5   &  8.832J & 8.237J & 7.994J & 0.28$\pm$0.06   & SB2  & no  & N/A   \\ 
LP~655-43ABC    &  12.91 &  14.44J & 13.14J & 11.41J &   1   &  9.730J & 9.136J & 8.818J & 0.18$\pm$0.04   & ST3  & yes & yes   \\ 
LTT~11586AcB    &  10.69 &  11.80J & 10.69J &  9.32J &   3   &  8.023J & 7.446J & 7.178J & 0.27$\pm$0.03   & ST2  & no  & yes   \\ 
LHS~1817Ab      &  12.30 &  13.69  & 12.50  & 10.84  &   4   &  9.096  & 8.464  & 8.176  & 0.24$\pm$0.02   & SB1  & no  & yes   \\ 
GJ~268AB        &   9.93 &  11.42J & 10.16J &  8.45J &   4   &  6.731J & 6.152J & 5.846J & 0.32$\pm$0.02   & SB2  & no  & no    \\
2MA~0930+0227AB &  12.40 &  \nodata& \nodata& \nodata& \nodata& 9.415J & 8.856J & 8.578J & 0.19$\pm$0.03   & SB2  & yes & yes   \\ 
LP~734-34AB     &  12.32 &  13.83J & 12.52J & 10.85J &   6   &  9.292J & 8.684J & 8.412J & 0.21$\pm$0.04   & SB2  & yes & yes   \\ 
G~123-45Ab      &  12.26 &  13.80 &  12.45  & 10.77 &    2   &  9.113  & 8.542  & 8.261  & 0.18$\pm$0.02   & SB1  & yes & yes   \\ 
GJ~1182AB       &  12.67 &  14.30J & 12.95J & 11.09J &   4   &  9.433J & 8.936J & 8.618J & 0.18$\pm$0.02   & SB2  & no  & no    \\ 
G~258-17AB      &  13.34 & \nodata &\nodata &\nodata &\nodata& 10.275J & 9.706J & 9.442J & 0.12$\pm$0.02   & SB2  & yes & yes   \\ 
LTT~7077AB      &  11.27 & 12.72J  & 11.44J & 9.78J  &   6   &  8.198J & 7.693J & 7.353J & 0.35$\pm$0.02   & SB2  & no & yes   \\
\enddata

\tablecomments{`J' indicates that the photometry is joint and
  therefore includes light from one or more companions. `SB1':
  single-lined binary, `SB2': double-lined binary, `ST2': double-lined
  triple, `ST3': triple-lined triple. {\it Gaia} $G$ photometry from \citet{GaiaDR2(2018)}; 2MASS $JHK$
photometry from \citet{Skrutskie(2006)}.}

\tablerefs{
(1) \citet{Reid(2002)};
(2)  RECONS, in prep (Silverstein et al.);
(3)  \citet{Weis(1991b)};
(4) \citet{Weis(1996)};
(5) \citet{Weis(1999)};
(6) \citet{Winters(2015)}.}

\end{deluxetable*}

\section{Data Acquisition} \label{sec:data}

We observed each object between UT 2016 October 14 and 2020 January 08
using TRES on the FLWO 1.5m Tillinghast Reflector. TRES is a
high-throughput, cross-dispersed, fiber-fed, echelle spectrograph. We
used the medium fiber ($2\farcs3$ diameter) for a resolving power of
$R \simeq 44\,000$. The spectral resolution of the instrumental
profile is 6.7 km s$^{-1}$ at the center of all echelle orders. For
calibration purposes, we acquired a thorium-argon hollow-cathode lamp
spectrum through the science fiber both before and after every science
spectrum. Exposure times ranged from $120$s to $3$~$\times$~$1200$s in
good conditions, achieving a signal-to-noise ratio of 3-25 per pixel
at $7150\ {\rm \AA}$ (the pixel scale at this wavelength is
$0.059\ {\rm \AA ~pix^{-1}}$). These exposure times were increased
where necessary in poor conditions. The spectra were extracted and
processed using the standard TRES pipeline \citep{Buchhave(2010)}.

We describe here the temperature and pressure control of
  TRES. TRES has two stages of temperature control.  The spectrograph
  is housed inside a custom enclosure with fine temperature control,
  which in turn is located in the coude room, which has coarse
  temperature control.  The temperature is monitored at more than a
  dozen key locations and the values are reported in the header of
  each observation.  Over several hours the variation is typically
  less than about 0.1 K at the spectrograph bench and echelle grating,
  with slow drifts from season to season of about 1 K.  There is no
  pressure control, so all science observations are sandwiched between
  wavelength calibrations using a thorium-argon hollow-cathode
  lamp. Drifts in the zero point of the velocity system are monitored
  using nightly observations of well-established radial-velocity
  standard stars, typically 3 or 4 per night.  These procedures are
  able to correct for drifts in the zero point of the velocity system
  to better than 10 m \pers ~from month to month, and better than about
  30 m \pers ~since 2013.  Orbital solutions for bright slowly-rotating
  stars typically have rms velocity residuals of 20 m \pers, which is
  a good indicator of the single-measurement precision when photon
  noise does not set the limit.

\section{Spectroscopic Analysis \& Radial Velocity Measurements} \label{sec:rv_analysis}

We use a template spectrum of Barnard's Star, observed on UT 2018 July
19, to perform cross-correlations based on the methods described in
\citet{Kurtz(1998)}. Barnard's Star is a slowly rotating \citep[130.4
  days,][]{Benedict(1998)} M4.0 dwarf \citep{Kirkpatrick(1991)} for
which we adopt a Barycentric radial velocity of $-110.3\pm0.5\ {\rm
  km\ s^{-1} }$, derived from presently unpublished CfA Digital
Speedometer \citep{Latham(2002)} measurements taken over 17 years. We
see negligible rotational broadening in our Barnard's Star template,
in agreement with the $v \sin i$ of 0.07 km \pers ~expected from the
130.4-day photometric rotation period noted above. This is also
consistent with the $v \sin i$ upper limit of $2 ~{\rm km\ s^{-1} }$
reported by \citet{Reiners(2018)}. We use the wavelength range 7065 to
7165\AA ~(echelle aperture 41) for the correlations, a region that is
dominated by TiO bandhead features in mid-to-late M dwarfs, which
provide many lines for the radial velocity (RV) measurements
\citep{Irwin(2011b)}. Part of the red end of the aperture is not
included, as it is contaminated by telluric absorption features.

For the double- and triple-lined systems, we use a least-squares
deconvolution \citep[LSD;][]{Donati(1997)} method to identify double-
or multi-lined systems and to estimate the initial light ratios
between the components. We then used these as starting points for {\sc
  todcor} and {\sc tricor} \citep{Zucker(1994),Zucker(1995)},
  which calculate the RV of each component in the double- and
  triple-lined systems, respectively. As with the single-lined
  systems, we used our observed Barnard's Star spectrum as the
  template for all components. In each case, we search for the light
ratio with the maximum correlation peak in echelle aperture 41, which
we then uniformly apply to all observations of each system to measure
the velocities. This assumes that there are no significant photometric
or spectroscopic variations in any of the stars in the system.

Aside from LHS~1817A, LTT~11586B, and 2MA~0930AB, none of the
  primary stars or components show any appreciable rotational
broadening at the resolution of the TRES spectra, so there was no need
to broaden the template spectrum in order to obtain a good
match. Therefore, we assumed a $v \sin i$ of zero for all systems
presented here, except for the three noted above. Because the
  light ratio is dependent upon the $v \sin i$ in the double-lined
  systems, we chose ever finer grids of $v \sin i$, for which we
  determined the light ratio that produced the maximum correlation
  peak via {\sc todcor}. We then fit preliminary orbits based on each
  set of resulting RVs. We chose the $v \sin i$ and light ratio that
  resulted in an orbital fit to the RVs with the smallest
  residuals. For the three systems noted above, the projected
  rotational velocities used to calculate the RVs are indicated in the
  the notes on those systems.

We use the Wilson method \citep{Wilson(1941)} to estimate the initial
mass ratio q $=$ m$_{2}$/m$_{1}$ and gamma velocity of the system as
inputs for the orbital fit. In short, we plot the velocities of the
primary component as a function of the velocities of the secondary
component. The negative slope of a linear fit to the velocities
provides the mass ratio, and the y-intercept divided by ($1 +
  q$) provides the gamma velocity.

Variations in the instrumental line profile on the order of 0.1 km
s$^{-1}$ result in systematic errors on components' velocities that
render them unreliable when their velocity difference is less than
roughly 2 km s$^{-1}$. We thus adopt 3.4 km s$^{-1}$, i.e., half the
spectral resolution of TRES, as a conservative upper limit to the
components' velocity difference and omit data where it is less than
this.

For systems that had MEarth data, we searched for evidence of
eclipses, which we are able to rule out for LHS~1817 and G~123-45.

\subsection{Orbit determination}
\label{subsec:orbit}

To determine orbital parameters, we used the same method as
  previously reported in \citet{Irwin(2011b),Winters(2018)}. We fit a
standard eccentric Keplerian orbit to the radial velocities of the
appropriate components in each system using the {\sc emcee} package
\citep{Foreman-Mackey(2013)} to obtain samples from the posterior
probability density function of the parameters using Markov Chain
Monte Carlo (MCMC).

Estimation of velocity uncertainties from cross-correlation analysis
is a notoriously difficult problem, particularly in the case of
multiple lined systems, so instead, and for consistency across all the
orbital solutions regardless of the number of spectroscopic
components, we take a simpler approach and derive the appropriate
velocity uncertainties $\sigma$ during model fitting.  Separate
uncertainties were allowed for each system component because these can
sometimes differ substantially, for example in SB2 systems with light
ratios very far from unity or with differing amounts of rotational
broadening.

In order to reduce the influence of lower quality spectra on the
results, the data points were weighted using the square of the
normalized peak cross correlation ($h_i$, as defined by
  \citet{Tonry(1979)}), where the resulting weights are $1/h_i^2$ for
data point $i$.  This is equivalent to adopting a velocity uncertainty
of $\sigma/h_i$ on data point $i$.  These resulting values are
reported in Table \ref{tab:rv-data}, but it is important to note that
they are not conventional, independent velocity uncertainties but are
rather derived from the orbit model and are dependent on the
assumption that the orbit model is the correct description of the
data. We note that we do not include the 0.5 km s$^{-1}$ uncertainty
on Barnard's Star's RV when fitting the orbit for each system. This is
because the orbit-fitting depends only on the relative velocity in all
parameters except the gamma velocity, and including this uncertainty
would cause the total uncertainties to be severely overestimated.

In practice, each solution must be validated by comparing the derived
value of $\sigma$ to our expectations based on simulated data or
experience from analysis of similar data sets.  The standard test of
comparing the $\chi^2$ value to the number of degrees of freedom as a
metric for goodness of fit is rendered useless by fitting for the
uncertainty (such solutions always produce this value of $\chi^2$, by
construction).

The resulting model has 7 free parameters for single-lined orbits, and
9 for double-lined orbits.  Five of these are common to both cases:
the epoch of inferior conjunction $T_0$, the orbital period $P$,
systemic radial velocity $\gamma$, $e \cos \omega$, and $e \sin
\omega$, where $e$ is eccentricity and $\omega$ is the argument of
periastron.  For single-lined orbits the remaining pair of parameters
are the velocity semi-amplitude $K$ and velocity uncertainty
$\sigma$. We calculate masses for the primary components of the
single-lined systems using the $K_{s}$ magnitude in Table
\ref{tab:phot} and the updated parallax $\pi_{DR2}$ listed in Table
\ref{tab:sample}. For double-lined orbits the remaining four
parameters are the mass ratio $q$, the sum of the velocity
semi-amplitudes of the two components $K_1+K_2$, and the separate
velocity uncertainties for each component $\sigma_1$ and $\sigma_2$.

This procedure results in a likelihood for the single-lined orbit
model (closely following the derivation in
\citealt{Gregory(2005)}) of:
\begin{equation}
p(D|M) = {\exp\left[-\sum_{i=1}^{N} {h_i^2 (v_i -
  m(t_i))^2\over{2 \sigma^2}}\right]\over{
  \sqrt{\prod_{i=1}^{N} 2 \pi \left({\sigma^2\over{h_i^2}}\right)}}}
\label{likelihood}
\end{equation}
where $D$ denotes the data and $M$ the model, $v_i$ are the individual
radial velocities for each data point $i$ (with $N$ data points
total), $h_i$ are their normalized peak cross-correlation values, and
$m(t_i)$ is the Keplerian model evaluated at data point $i$ (time
$t_i$).  The product of the normalization constants of the Gaussian
distributions for the data points appearing in the denominator of
Eq. (\ref{likelihood}) is often omitted but must be included
explicitly when fitting for $\sigma$ because it is no longer constant.
In practice, we calculate $\log p(D|M)$ when implementing this method
to avoid the exponential and product in Eq. (\ref{likelihood}) but we
have given the equation for the likelihood itself for clarity.

For double-lined orbits each observation results in a pair of
velocities, and these were treated as two data points using the same
likelihood formulation, but substituting the appropriate $\sigma_1$ or
$\sigma_2$ parameters as needed for each component to replace $\sigma$
in Eq. (\ref{likelihood}) and calculating the model velocity for the
respective component in place of $m$.

For the $\sigma$ parameters we adopt modified Jeffreys priors of the
form:
\begin{equation}
p(\sigma) \propto {1\over{\sigma + \sigma_a}}
\end{equation}
where $\sigma > 0$ and $\sigma_a > 0$ is a constant.  The value of
$\sigma_a$ was set to $10\%$ of the estimated velocity uncertainty,
following \citet{Gregory(2005)}, Eq. (16) and surrounding
discussion.

The choice of $e \sin \omega$ and $e \cos \omega$ as jump parameters
was made for mathematical convenience but has the undesirable feature
of producing a linearly increasing prior on $e$ if uniform priors are
adopted on these two parameters.  In order to avoid this we include a
factor of $1 / e$ in the prior and reject any points with $e \ge 1$ to
convert this to a uniform prior on $e \in [0,1)$.

Uniform improper priors were adopted on all other parameters,
resulting in an overall prior probability density function
proportional to the product of the factors from Eq. (2) and $1/e$
from the previous paragraph.  This was then combined with the $\log$
likelihood from above to obtain the $\log$ posterior returned by the
objective function to {\sc emcee}.

The MCMC simulations were initialized using a Levernberg-Marquardt
(L-M) fit of the function $m$ for the Keplerian parameters, and the
rms of the residuals of the data about this model were used to
initialize the $\sigma$ parameters.  One hundred walkers, as
  described by the {\sc emcee} method, were then populated with
initial values derived by perturbing the L-M fit results by Gaussian
random deviates with a standard deviation of $3$ times the estimated
parameter uncertainties from the L-M to ensure different starting
points for each walker and burned in for $1 \times 10^4$ samples,
followed by $5 \times 10^4$ samples of the posterior probability
density function retained for analysis.  The samples from all of the
walkers were combined assuming independence resulting in a total of $5
\times 10^6$ samples, which were converted to a central value and
uncertainty using the median and $68.3$ percentile of the absolute
deviation from the median, respectively.

This paper also includes orbits for two more complicated triple
systems.  For these solutions we simply extended the model already
described above to allow for a variable $\gamma$ velocity, but
otherwise ignored the additional component during fitting and MCMC
analysis.  These additional (outer) components have incomplete orbits
and large period ratios so these models simply used linear or
quadratic functions for the $\gamma$ velocity.

We describe each system individually here, ordered from single- to
triple-lined systems and by RA within each section. The measured RVs
for each system, along with their internal uncertainties, are reported
in Table \ref{tab:rv-data}. We show the orbital fits to the TRES RVs
(top) and residuals to the fits (bottom) in Figures
\ref{fig:lhs1817_orbit} -- \ref{fig:ltt07077_lp655_orbit}, with
primaries shown in blue, secondaries in green, and tertiaries in
red. Error bars are smaller than the points in the orbital fit plots,
in most cases, and are repeated in the residuals plots. The orbital
parameters for GJ~268AB are listed in Table \ref{tab:gj0268}, and we
present the orbital parameters for all other systems in Table
\ref{tab:orb_el}, ordered by RA. The only exception is L~870-44AB; no
orbit has been derived yet for this system because there are not yet
enough data for a robust period determination. Finally, we assign
component designations with capital letters if we see the spectral
lines and lower case letters if we do not see the lines; for example,
we use `Ab' for the single-lined binaries and `AB' for the
double-lined binaries. For the triple systems, the component
indicators are assigned according to the height of the peak of each
component in the least squares deconvolution analysis, where the
primary component has the largest peak and is assigned the designation
`A'.

\subsection{Single-lined Systems}
\label{subsec:singles}

We describe the two single-lined systems, LHS~1817Ab and G~123-45Ab.

\subsubsection{LHS~1817Ab}
\label{subsubsec:lhs1817}

\begin{figure*}
\minipage{0.50\textwidth}
\includegraphics[scale=.45,angle=0]{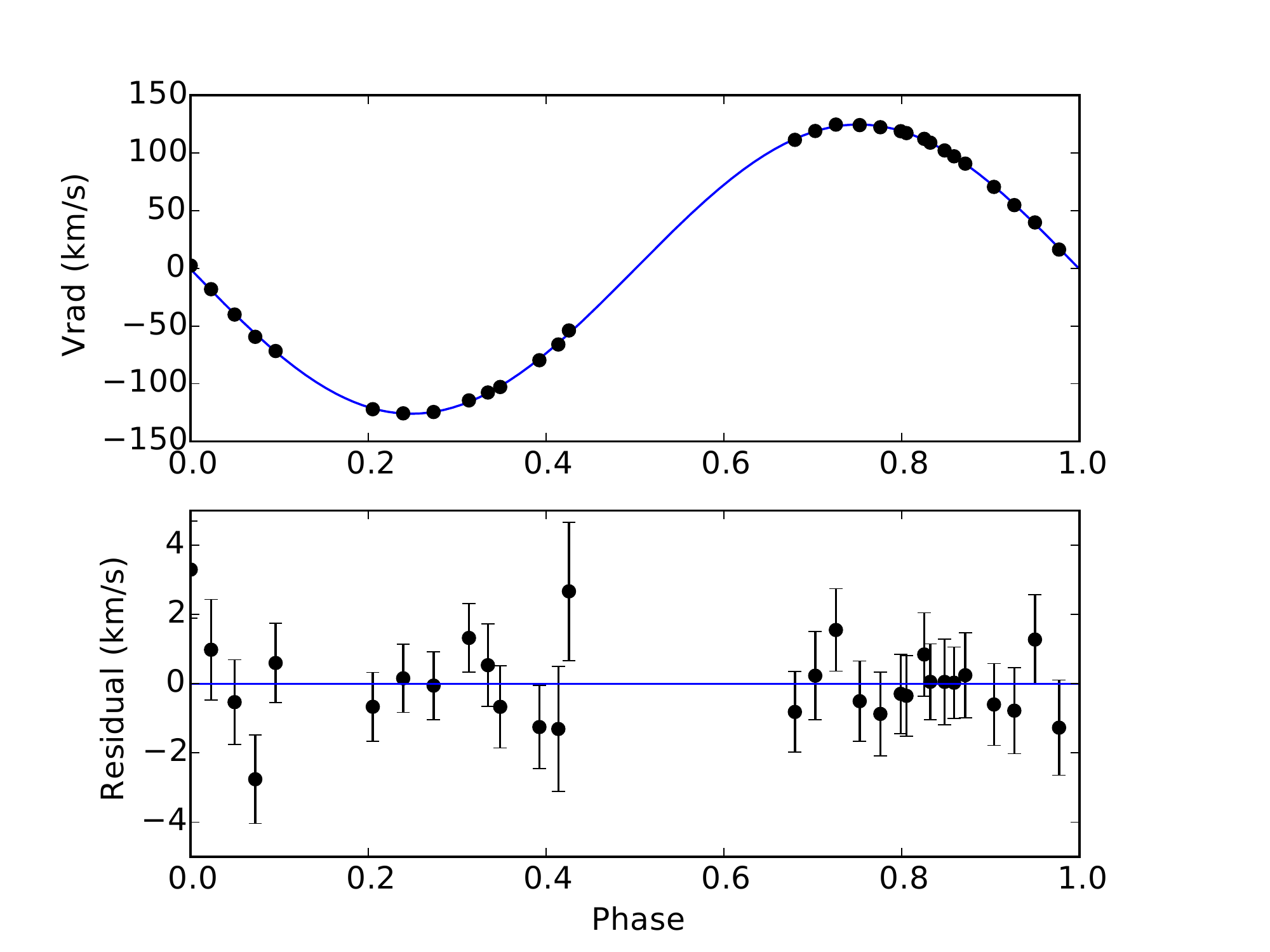}
\endminipage\hfill
\hspace{-0.3cm}
\minipage{0.50\textwidth}
\includegraphics[scale=.38,angle=270]{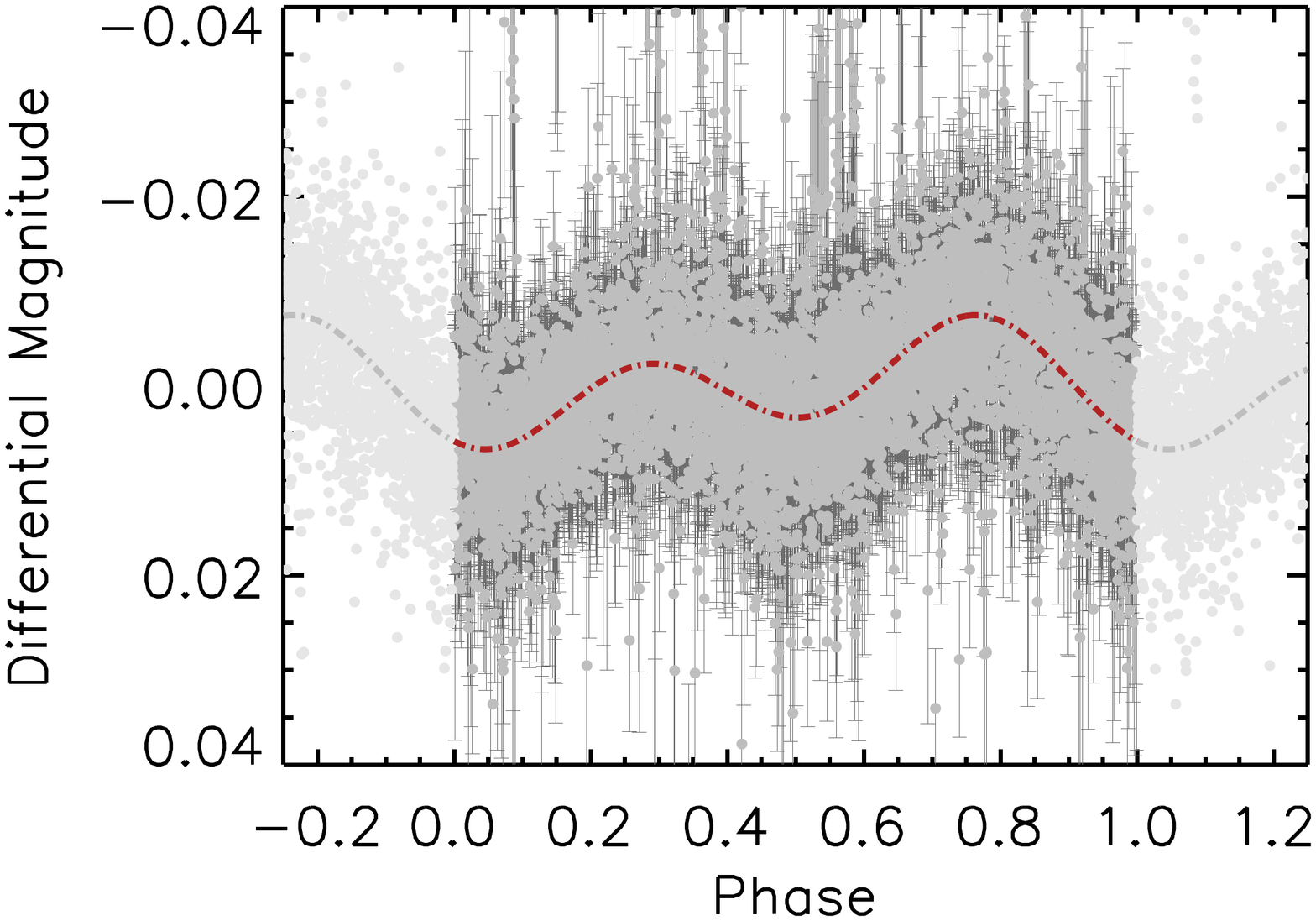}
\endminipage\hfill
\caption{The orbital fit (left) and light curve (right) of
  LHS~1817A. ({\it Left}) The best orbital fit to the velocities is
  shown in the top panel, while residuals to the fit are indicated in
  the bottom panel. ({\it Right}) The MEarth light curve of LHS~1817A,
  phase-folded to the orbital (and rotational) period of $0.3099267
  \pm 0.0000014$ days. The dash-dot red line indicates the fit
  that includes the ellipsoidal modulation due to the white dwarf. The
  lighter shaded regions at phases -0.25-0.0 and 1.0-1.25} are
duplicated portions of the light curve. \label{fig:lhs1817_orbit}
\end{figure*}

This system is a single-lined, spectroscopic binary that was suggested
by \citet{Newton(2016)} to be multiple because of an unphysically
metal-rich estimate of the star's metallicity. We obtained 24
observations of LHS~1817A with S/N per spectral resolution element of
5-18. The typical data processing of the spectra of our targets
involves median combining the three exposures usually acquired to
minimize the effects of radiation events. However, the combined
spectra of this target result in exposure times of 900 - 2700s and
comprise a significant fraction of the 7.4-hour orbital period
(roughly 3 -- 10\%), resulting in the velocities smearing. Therefore,
we analyze the individual spectra from each set of exposures to
calculate the velocities. This generally results in lower, but still
usable, correlation function peak heights. We use the mean $v \sin i$
of $27.9\pm1.3$ km s$^{-1}$ from all epochs, in agreement with that of
$29.7\pm2.6$ km s$^{-1}$ reported by \citet{Kesseli(2018)}, to
calculate the velocities. We estimate a mass of $0.290 \pm 0.014$
M/M$_{\odot}$ for the primary star. We find an orbital period of
$0.30992678 \pm 0.00000048$ days, an eccentricity of $0.0063 \pm
0.0031$, and a minimum mass ratio of $0.9343 \pm 0.0035$. Our measured
orbital period is in agreement with the photometric rotation period
reported by \citet{Newton(2016)} and indicates that the system has
synchronized, as expected due to the short orbital period. We show the
orbital fit and velocity residuals in the left panel of Figure
\ref{fig:lhs1817_orbit}.

The resulting minimum mass ratio from our orbital fit is near
unity, but we do not see a second set of spectral lines from the
companion. We therefore expect that the companion is a white dwarf. We
have five years of MEarth \citep{Nutzman(2008),Irwin(2015)} data for
this target, taken UT 2011 October 11 -- 2016 October 30. The light
curve shows two peaks, an indication of ellipsoidal modulation, where
a massive companion is tidally distorting the M dwarf into an
ellipsoidal shape. This lends strength to our assumption that the
companion is a white dwarf. We modeled the ellipsoidal modulation by
fitting to the light curve a sine curve plus a second harmonic
model. We show the light curve with our fit to the ellipsoidal
variation in the right panel of Figure \ref{fig:lhs1817_orbit},
phase-folded to the photometric rotation period.

The amplitude of the ellipsoidal variation depends on inclination as
$sin^2 i$, the ratio of the stellar radius to the semi-major axis
$R_1/a$, the mass ratio $q$, and limb darkening and gravity darkening
(e.g., \citealt{Shporer(2017)}).  The dependence on inclination is
different from that of the semi-amplitude $K$ as a function of the
same parameters, so it is possible to infer the inclination by
combining these constraints.

In order to do so, we estimate the stellar radius using the
mass-luminosity and mass-radius relations, and obtain limb darkening
and gravity darkening coefficients from the tabulations of
\citet{Claret(2012)}, which require the effective temperature.

We estimate the radius of the M dwarf to be $0.293\pm0.027$
R$_{\odot}$ using the empirical single star mass-radius relation in
\citet{Boyajian(2012)}, based on the mass of the star ($0.290\pm0.014$
M$_{\odot}$) calculated using the MLR by \citet{Benedict(2016)}. We
note that these relations may not be entirely appropriate for this
system, as the M dwarf component likely accreted material from the
white dwarf's progenitor when it evolved off the main sequence. We
also find it to be slightly overluminous in the $V-$ and $K-$bands;
thus, it falls among the blended photometry binary sequence on an
observational Hertzsprung-Russel diagram. We note that the system is
X-ray-bright, as previously reported by \citet{Shkolnik(2012)}.

We estimate the effective temperature by combining two relations.
Using the Stefan-Boltzmann Law we obtain a value of $3333\pm150$ K,
using the method previously described by
\citet{Dittmann(2017a),Ment(2019)}. From the ($V - J$), ($J - H$)
relation in \citet{Mann(2015),Mann(2016)}, we calculate the effective
temperature to be $3207\pm77$ K.  We adopt the unweighted mean of these
two values $3270\pm150$ K.

To infer the inclination, we assume all the signal seen in the second
harmonic in the light curve analysis is due to ellipsoidal variation,
ignoring the fundamental, which seems to be spot-dominated.  While the
second harmonic will inevitably be somewhat polluted by spots, the
phase being fairly consistent with that of the orbital solution
implies that most of the signal is due to the ellipsoidal modulation.

We use an MCMC analysis to obtain the posterior for the inclination,
including the mass-luminosity and mass-radius relations, limb
darkening and gravity darkening table lookups inside this procedure to
allow for the uncertainties.  Gaussian priors were adopted for the
orbital period, second harmonic semi-amplitude, effective temperature,
K magnitude, and parallax derived from the observations as appropriate
and the resulting posterior for the inclination analysed as described
in \S \ref{subsec:orbit}.

The resulting estimate for the orbital inclination is
27.8$\pm$0.96\arcdeg. This yields a mass for the white dwarf of
1.03$\pm$0.08 M$_{\odot}$ and a mass ratio of $3.57\pm0.14$ for the
system.

As a comparison, we can also independently estimate the inclination if
we assume that the rotational axis of the M dwarf is aligned with the
orbital axis of the binary system. Because we know the photometric
rotation period $P_{rot}$ of the M dwarf, as well as its $v \sin i$,
we use the relation $P_{rot} * v \sin i = 2 \pi R \sin i$. We note
that the magnitude of the $v \sin i$ may not be due entirely to
rotational broadening and may have an additional broadening
contribution from other sources for which we have not accounted. From
these admittedly imperfect assumptions and adopting ten percent errors
for the rotation period, we estimate the inclination of the system to
be $36.3\pm6.3$\arcdeg. From this inclination, we calculate a mass of
$0.64^{+0.25}_{-0.14}$ M$_{\odot}$ for the white dwarf, which results
in a mass ratio of $2.21^{+0.87}_{-0.50}$ for the system.

\begin{deluxetable}{lrrcr}
\tabletypesize{\scriptsize}
\tablecaption{Radial Velocities for Multiple Systems \label{tab:rv-data}}
\tablecolumns{5}
\tablenum{3}

\tablehead{
\colhead{BJD\tablenotemark{a}} & 
\colhead{A $v_{\rm rad}$\tablenotemark{b}} &
\colhead{B $v_{\rm rad}$\tablenotemark{b}} &
\colhead{C $v_{\rm rad}$\tablenotemark{b}} &
\colhead{$h$\tablenotemark{c}} \\ 
\colhead{(days)} & 
\colhead{(${\rm km\ s^{-1}}$)} &
\colhead{(${\rm km\ s^{-1}}$)} &
\colhead{(${\rm km\ s^{-1}}$)} &
\colhead{} 
}
\startdata
\hline
GJ~1029AB    &         &         &          &            \\
\hline
2457759.6280 & -20.121$\pm$0.123 &   0.474$\pm$0.964 & \nodata& 0.802994 \\
2457944.9249 & -19.027$\pm$0.121 &   0.165$\pm$0.945 & \nodata& 0.819148 \\
2457993.9720 &  -6.797$\pm$0.122 & -17.107$\pm$0.952 & \nodata& 0.813055 \\
2458002.8908 &  -7.730$\pm$0.117 & -17.120$\pm$0.916 & \nodata& 0.845087 \\
2458027.8068 & -13.272$\pm$0.112 &  -9.713$\pm$0.874 & \nodata& 0.885454 \\
2458034.8799 & -16.250$\pm$0.109 &  -3.582$\pm$0.849 & \nodata& 0.912088 \\
2458042.7260 & -19.790$\pm$0.115 &   0.208$\pm$0.903 & \nodata& 0.857204 \\
2458050.7436 & -18.108$\pm$0.111 &  -1.706$\pm$0.870 & \nodata& 0.889733 \\
2458055.8456 & -13.808$\pm$0.109 &  -9.021$\pm$0.853 & \nodata& 0.907916 \\
2458063.7882 &  -8.912$\pm$0.114 & -14.354$\pm$0.890 & \nodata& 0.869500 \\
2458082.6561 &  -6.668$\pm$0.113 & -18.136$\pm$0.880 & \nodata& 0.879921 \\
2458090.7040 &  -7.083$\pm$0.120 & -18.430$\pm$0.936 & \nodata& 0.826577 \\
2458107.7450 &  -8.910$\pm$0.116 & -13.698$\pm$0.909 & \nodata& 0.851294 \\
\enddata
\tablenotetext{a}{Barycentric Julian Date of mid-exposure, in the TDB
  time-system.}
\tablenotetext{b}{Barycentric radial velocity. The internal
  model-dependent uncertainties on each listed velocity are $\sigma$/$h$, where $\sigma$ is listed in Table \ref{tab:orb_el} and $h$ is the peak-normalized cross-correlation for each spectrum listed here. }
\tablenotetext{c}{Peak normalized cross-correlation.}
\tablecomments{The velocities for the first system in our sample are
  shown to illustrate the form and content of this table. The full
  electronic table is available in the online version of the paper.}

\end{deluxetable}

\subsubsection{G~123-45Ab}
\label{subsubsec:g123}

\begin{figure*}
\minipage{0.50\textwidth}
\includegraphics[scale=.45,angle=0]{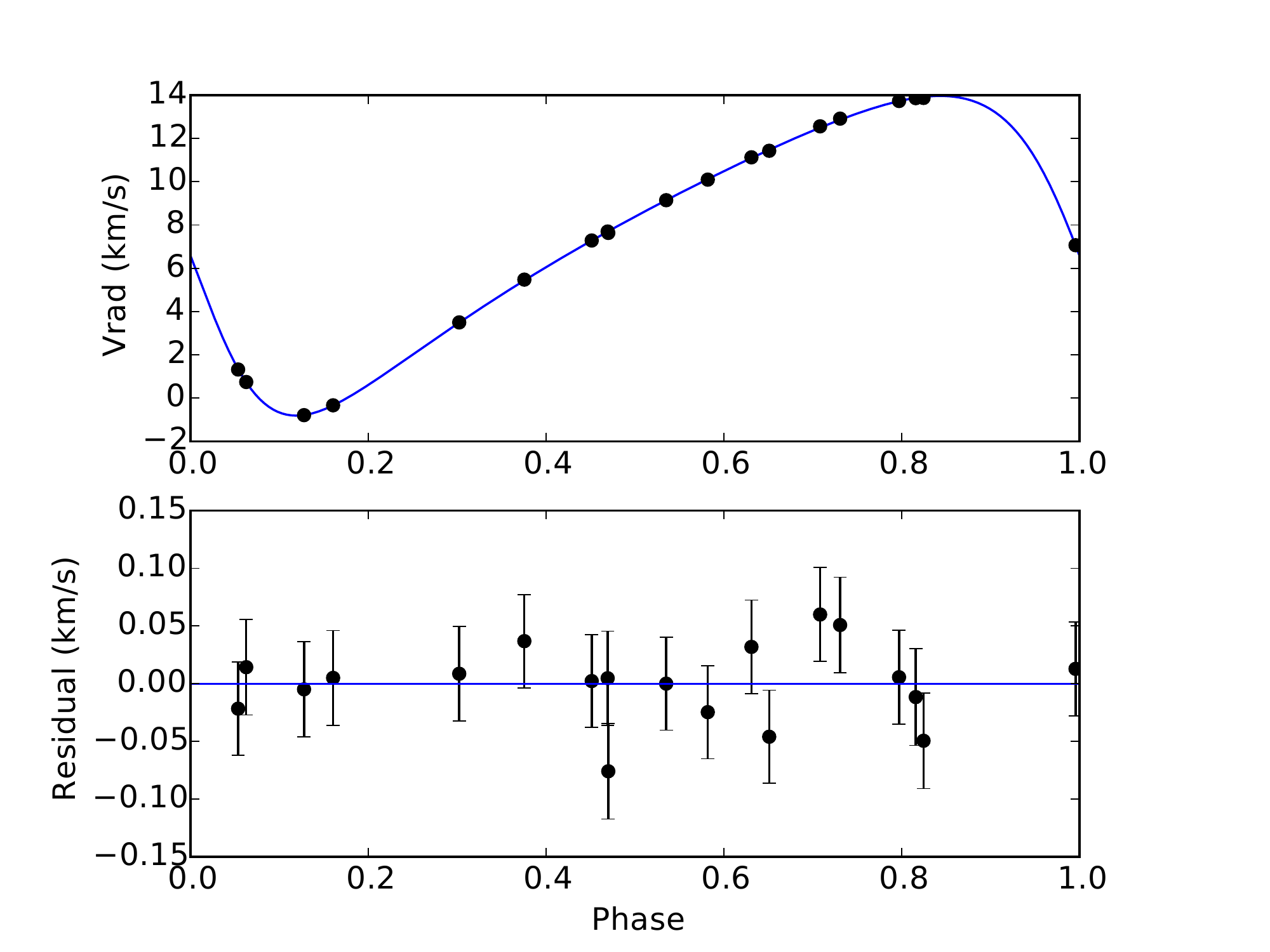}
\endminipage\hfill
\hspace{-0.3cm}
\minipage{0.50\textwidth}
\includegraphics[scale=.45,angle=0]{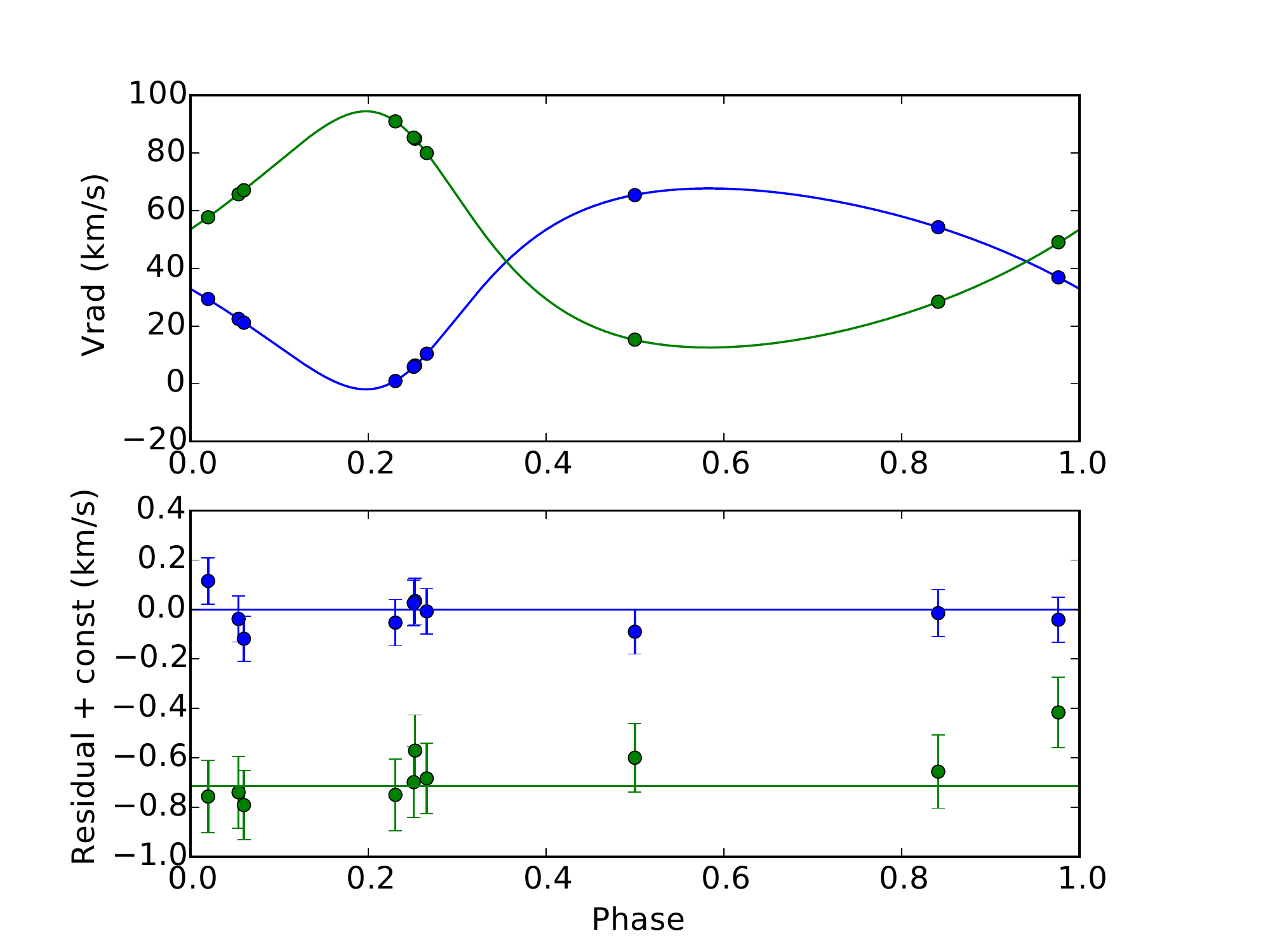}
\endminipage\hfill
\caption{The orbital fits of G~123-45A and GJ~268AB. The best orbital fits to the data for G~123-45A (left) and GJ~268AB (right) are shown in the top panels, while residuals to the fit are indicated in the bottom panels. \label{fig:g123_gj268_orbit}}
\end{figure*}

This is a new single-lined spectroscopic binary. We obtained 19
spectra with S/N per spectral resolution element of 19-39 and see no
evidence of broadening due to rotation. We estimate a mass of $0.185
\pm 0.014$ M/M$_{\odot}$ for the primary star. We find an orbital
period of $34.7557 \pm 0.0041$ days, an eccentricity of $0.3758 \pm
0.0024$, and a minimum mass ratio of $0.20903 \pm 0.00049$ for the
system. We measure the minimum mass of the companion to be $40.510 \pm
0.096$ M$_{J}$, significantly below the sub-stellar boundary. We show
the resulting orbital fit to the velocities in the left panel of
Figure \ref{fig:g123_gj268_orbit}.

\subsection{Double-lined Systems}
\label{subsec:doubles}

We present the orbit of GJ~268 to illustrate the robustness of our
method for double-lined binaries, followed by the results for the
  remaining systems.

\subsubsection{GJ~268AB}
\label{subsubsec:gj0268}

The orbit for this well-known, double-lined binary has been previously
published \citep{Tomkin(1986),Delfosse(1999c)}. We chose to measure
the orbit of this system in order to compare the results of our method
with previous results. We acquired 10 spectra with S/N per spectral
resolution element of 23-34. We see no evidence of rotational
broadening and use a light ratio of 0.692 to derive the velocities of
the components with {\sc todcor}. All the velocities were
well-separated, with none needing to be discarded. We show the results
of our analysis, along with those from \citet{Delfosse(1999c)} in
Table \ref{tab:gj0268}. There is excellent agreement between the two
orbital solutions, when taking into account the 0.5 km \pers ~RV
uncertainty of our Barnard's Star template. We present the orbital fit
for GJ~268AB in the right panel of Figure \ref{fig:g123_gj268_orbit}.

\begin{deluxetable*}{lcc}
\tabletypesize{\small}
\tablecaption{Comparison of GJ~268AB Orbital Parameters \label{tab:gj0268}}
\tablecolumns{3}
\tablenum{4}

\tablehead{
\colhead{Parameter}            & 
\colhead{This Work}            &
\colhead{\citet{Delfosse(1999c)}}
}

\startdata
\hline
MCMC parameters                       &                           &                                  \\
\hline
$e \cos \omega$                       & $-0.2721 \pm 0.0018$      & \nodata                     \\  
$e \sin \omega$                       & $-0.1736 \pm 0.0016$      & \nodata                     \\   
$T_0$ (BJD)                           & $2457675.4569 \pm 0.0064$ &  \nodata                    \\  
$P$ (days)                            & $10.42673 \pm 0.00010$    & $10.4265 \pm 0.00002$       \\  
$q$\tablenotemark{a}                  & $0.8505 \pm 0.0022$       & $0.851 \pm 0.001$    \\       
$\gamma$ (km s$^{-1}$)\tablenotemark{b} & $42.337 \pm 0.039$      & $41.83 \pm 0.03$             \\
$(K_1+K_2)$ (km s$^{-1}$)\tablenotemark{a} & $75.78 \pm 0.13$     & $75.67 \pm 0.07$ \\ 
$\sigma_1$ (km s$^{-1}$)                & $0.084 \pm 0.028$       & \nodata                     \\ 
$\sigma_2$ (km s$^{-1}$)                & $0.131 \pm 0.038$      & \nodata                     \\ 
\hline
Derived parameters                   &                           &                             \\
\hline
$e$                                  & $0.3227 \pm 0.0018$        & $0.321 \pm 0.001$           \\       
$\omega$ (deg)                       & $212.54 \pm 0.30$          & $212.1 \pm  0.3$            \\       
$a \sin i$ (AU)                      & $0.06873 \pm 0.00011$      & \nodata                     \\       
$(M_1+M_2) \sin^3 i$ (${\rm M}_\odot$)\tablenotemark{a} & $0.39854 \pm 0.00186$ & $0.398 \pm 0.001$ \\ 
$M_1 \sin^3 i$ (${\rm M}_\odot$)     & $0.21535 \pm 0.00110$      & $0.215 \pm 0.001$            \\    
$M_2 \sin^3 i$ (${\rm M}_\odot$)     & $0.18319 \pm 0.00081$      & $0.183 \pm 0.001$            \\    
$T_{\rm peri}$ (BJD)\tablenotemark{c} & $2457677.9835 \pm 0.0081$  & $2450149.902 \pm 0.008$  \\ 
${\alpha}_{ast}$ (mas)                 &  0.7                        & \nodata \\
\enddata

\tablenotetext{a}{The value listed here was not specifically reported in \citet{Delfosse(1999c)} and has been calculated.}
\tablenotetext{b}{We note that the uncertainty listed here is our internal uncertainty; when calculating the total uncertainty on the systemic velocity, one should add in quadrature the 0.5 km s$^{-1}$ uncertainty on the radial velocity of our template Barnard's Star.}
\tablenotetext{c}{The $T_0$ value reported in \citet{Delfosse(1999c)} is equivalent to our calculated $T_{\rm peri}$ value, where the difference is the 722 periods that have elapsed between the two measurements.}

\end{deluxetable*}

\begin{figure*}
\minipage{0.50\textwidth}
\includegraphics[scale=.45,angle=0]{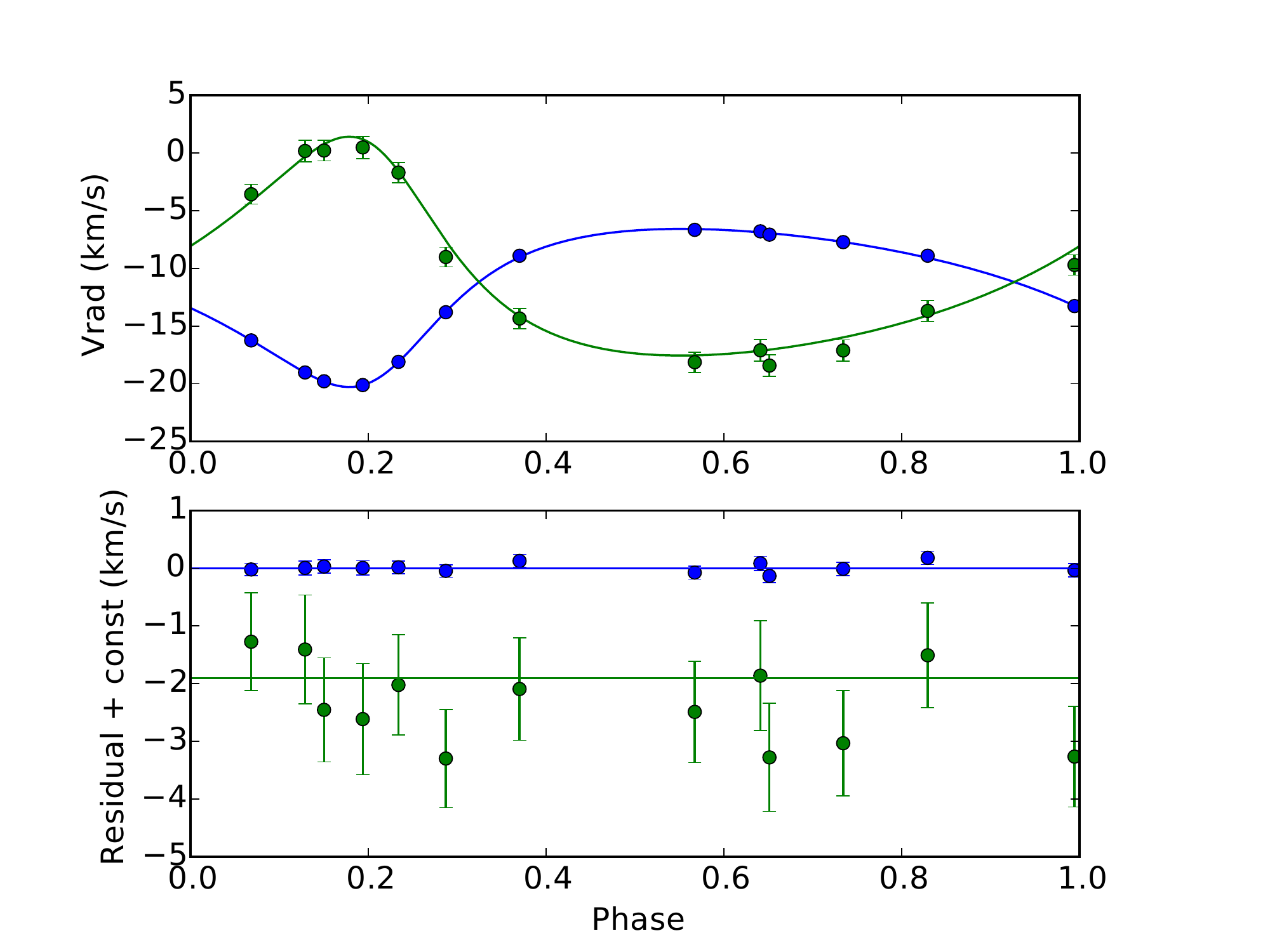}
\endminipage\hfill
\minipage{0.50\textwidth}
\includegraphics[scale=.45,angle=0]{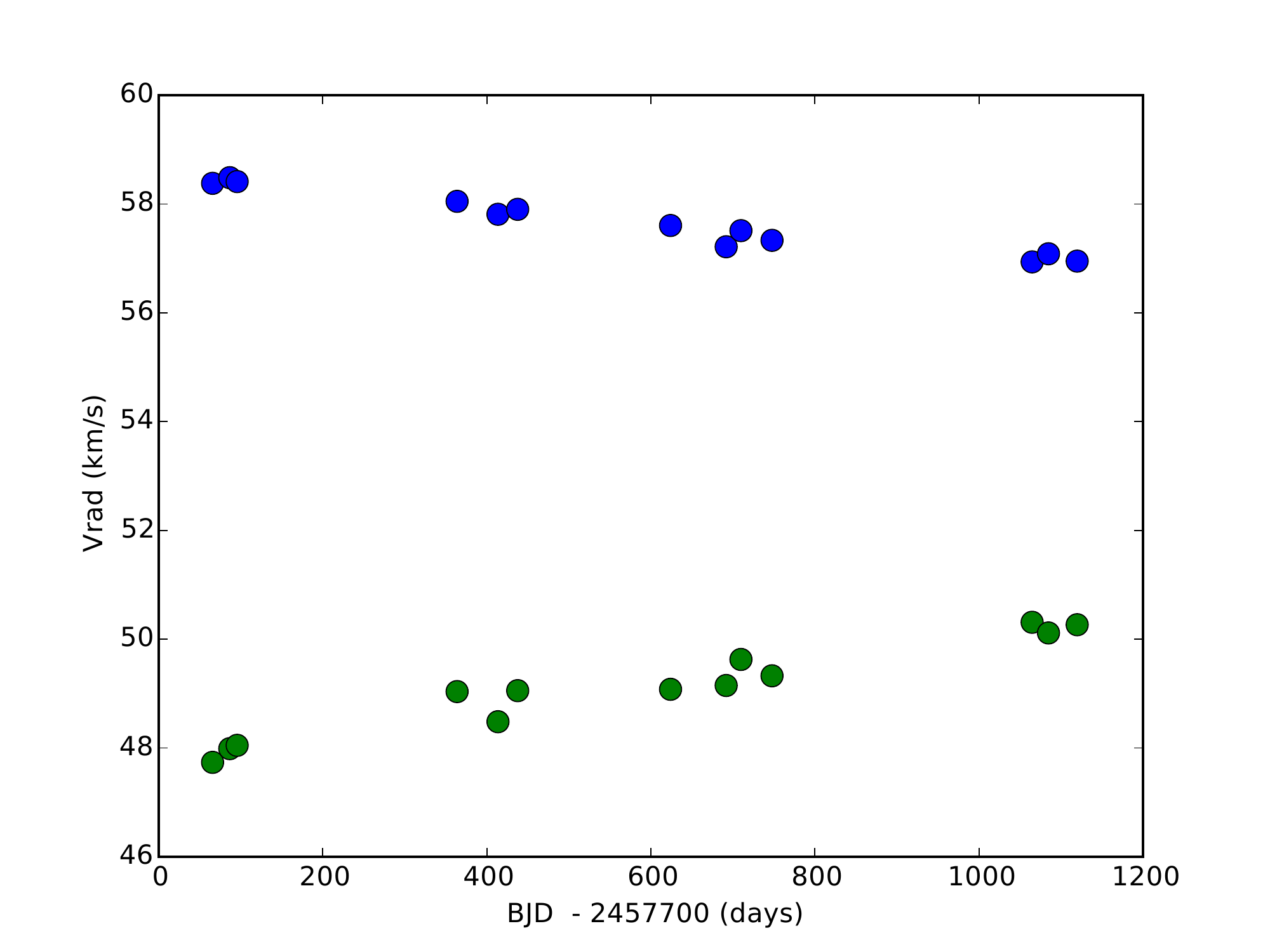}
\endminipage\hfill
\caption{The orbital fit of GJ~1029AB (left) and the radial velocities of L~870-44AB (right).  \label{fig:gj1029_l870_orbit}}
\end{figure*}

\subsubsection{GJ~1029AB}
\label{subsubsec:gj1029}

An orbit for this system was reported in \citet{Baroch(2018)}; we
independently detected the presence of doubled lines in 2017 and
proceeded to measure the orbit. We gathered 15 spectra with S/N per
spectral resolution element of 13-21. We see negligible rotational
broadening. We discarded two epochs with insufficient velocity
separation, and used a light ratio of 0.223 to calculate the
velocities. We find an orbital period of $95.76 \pm 0.18$ days, in
agreement with the period reported in \citet{Baroch(2018)}. We measure
an eccentricity of $0.3786 \pm 0.0067$ and a mass ratio of $0.722 \pm
0.026$. We show the resulting orbital fit in the right panel of Figure
\ref{fig:gj1029_l870_orbit}.

\subsubsection{L~870-44AB}
\label{subsubsec:l870}

This system was included in our initial sample with a photometric distance of 14.6$\pm$2.2 pc \citep{Winters(2015)}, but the {\it Gaia} DR2 reports a trigonometric distance of 25.6$\pm$0.7 pc. \citet{Jodar(2013)} reported a visual companion detected with lucky imaging to have an angular separation of 0\farcs238 at a position angle of 189\arcdeg ~in 2008 with a $\Delta$~$I$ of 1.37 mag between the components. 

We report the detection of double lines in the spectrum of this object
and a long-term trend in the velocities. With twelve spectra of S/N
per spectral resolution element of 16-34 taken over nearly three
years, we estimate the orbital period of this system to be roughly 19
years. This is in agreement with the roughly 18-year orbital period we
estimate from the lucky imaging results, assuming a circular orbit
with a semi-major axis equal to the angular separation. The flux ratio
of 0.369 derived from our {\sc TODCOR} analysis of the spectra near
710 nm is roughly in agreement with the reported magnitude difference
at $I$ in \citet{Jodar(2013)} (corresponding to a flux ratio of 0.28),
so we are confident that the component observed in our spectra is the
same. Neither component show detectable rotational broadening. Using
the Wilson method, we derive a mass ratio of $0.60\pm0.05$ and a gamma
velocity of $54.4\pm1.5$ km s$^{-1}$. We do not report an orbit for
this system; thus, we do not report uncertainties on the RVs. We show
the RVs of the components as a function of time in the right panel of
Figure \ref{fig:gj1029_l870_orbit}, and we include our measured radial
velocities, without uncertainties, in Table \ref{tab:rv-data}.

\begin{figure*}
\minipage{0.50\textwidth}
\includegraphics[scale=.47,angle=0]{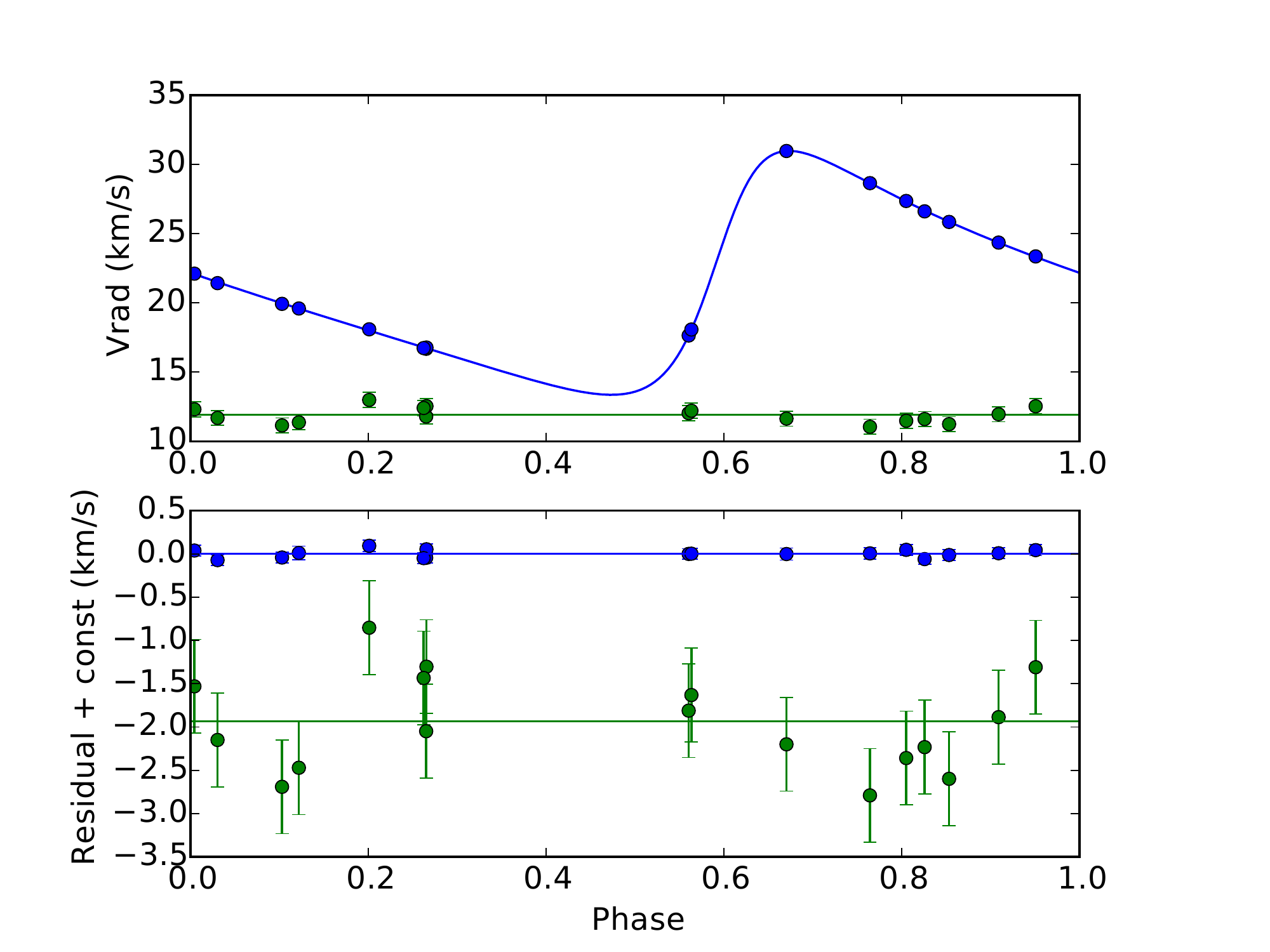}
\endminipage\hfill
\minipage{0.50\textwidth}
\includegraphics[scale=.47,angle=0]{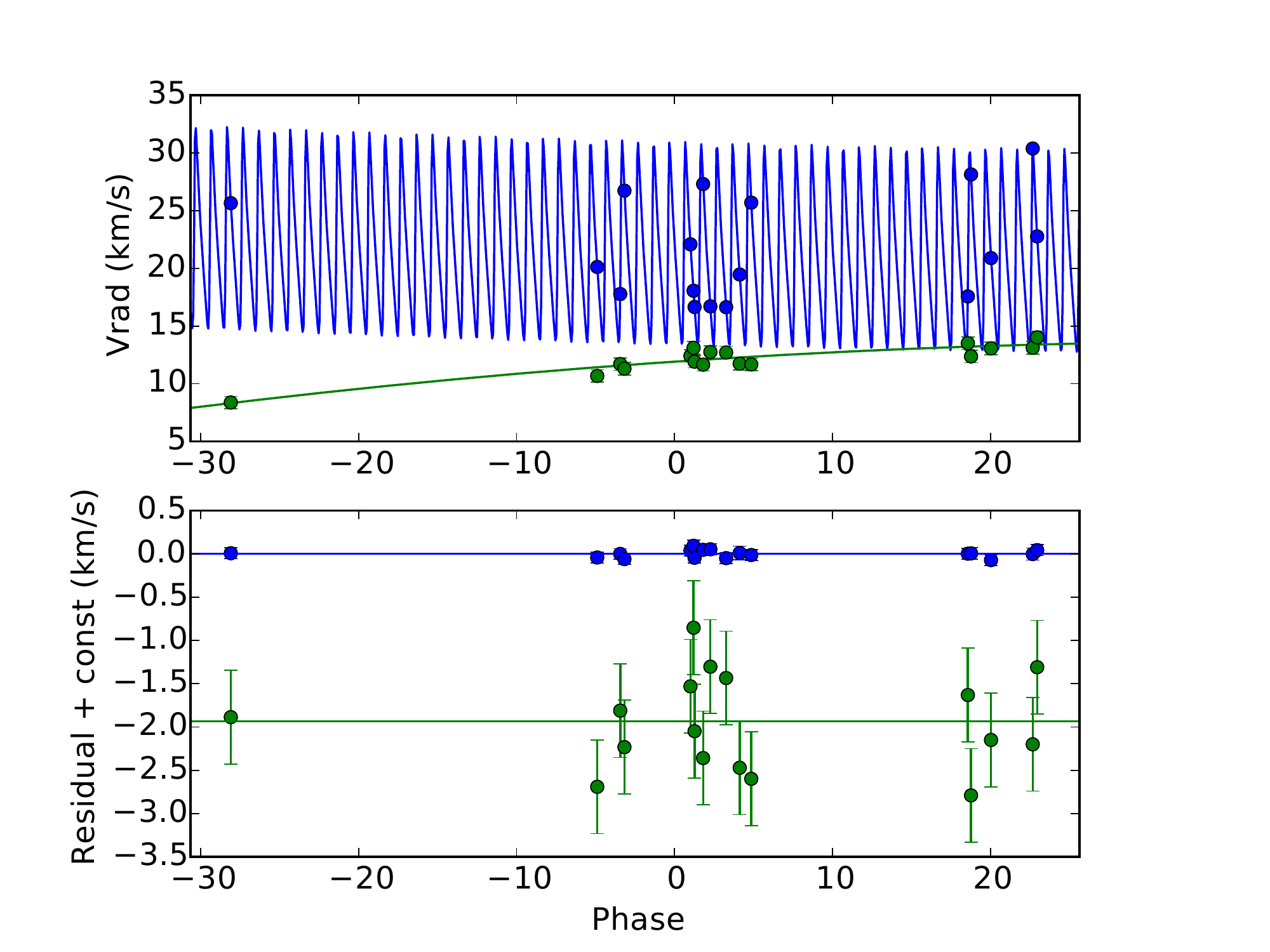}
\endminipage\hfill
\caption{The folded (left) and unfolded (right) orbital fits of LTT~11586AcB, where the unfolded orbit illustrates the velocity drift, modeled as a quadratic, of the outer `B' component. \label{fig:ltt11586_orbits}}
\end{figure*}

\subsubsection{LTT~11586AcB}
\label{subsubsec:ltt11586}

This is a known triple system, which is reported to be an SB1
\citep{Jeffers(2018)} with a visual component resolved with lucky
imaging \citep{Cortes-Contreras(2017)} with a $\Delta$~$I_{\rm KC}$ of
$1.96 \pm 0.05$ mag at a separation of $0\farcs540 \pm 0\farcs003$. We
confirm that it is a triple system where we see doubled spectral lines
and present the orbit of the inner SB1 here. We refer to the primary
component as `A', the widely separated component, whose spectral lines
we detect, as `B', and the primary's unseen companion as `c'.

We obtained 20 spectra of this system with S/N per spectral resolution
element of 15-41. We estimate masses for A and B by converting the
$\Delta$~$I_{\rm KC}$ to $\Delta$~$K_{\rm s}$ using the relations in
\citet{Riedel(2014)} and then using the MLR by \citet{Benedict(2016)}.
We find masses for A and B of $0.289 \pm 0.017$ and $0.134 \pm 0.014$
M/M$_{\odot}$. We determine a light ratio of 0.202 between A and B and
use a rotational velocity of 4 km s$^{-1}$ for the B to calculate the
velocities of A and B, constraining the velocities of B to be between
8 and 15 km s$^{-1}$. We fit a quadratic drift ($\ddot{\gamma}$) to B
and discarded three observations with insufficient velocity separation
between A and B. We find an orbital period of $15.04547 \pm 0.00041$
days, an eccentricity of $0.5029 \pm 0.0055$, and a minimum mass ratio
of $0.1462 \pm 0.0014$ for the SB1. The unseen component of the SB1,
c, for which we measure a minimum mass of $44.27 \pm 0.44$ M$_J$, is
significantly below the stellar-substellar boundary. We show the orbit
in the left panel of Figure \ref{fig:ltt11586_orbits}. We also show
the unfolded orbit in the right panel to illustrate the velocity drift
of the outer B component in the system. We do not report velocity
uncertainties for the B component; the uncertainties shown in the
orbital fits for this component are the rms of the velocities.

We calculate a mass ratio of $0.40 \pm 0.02$ for B/Ac. Assuming a
circular orbit and a semi-major axis between Ac and B equal to the
reported angular separation in \citet{Cortes-Contreras(2017)}, we
estimate an orbital period for B around Ac of 23 years.

\subsubsection{2MA~0930+0227AB}
\label{subsubsec:2ma0930}

This new double-lined spectroscopic binary had no existing spectrum in
the literature when it was added to our target list. We detected a
faint secondary that had significant rotation and gathered 25 spectra
of the system with S/N per spectral resolution element of 15-28. We
adopted a light ratio of 0.404 to calculate the radial velocities with
rotational velocities of 3 and 26 km \pers ~for the primary and
secondary components, respectively. We determine an orbital period of
$916.8 \pm 2.5$ days, an eccentricity of $0.1928 \pm 0.0046$, and a
mass ratio of $0.667 \pm 0.035$ for 2MA~0930+0227AB. We show the
orbital solution to the TRES data in the left panel of Figure
\ref{fig:2ma0930_lp734_orbit}. We note that the residuals to the fit
of the B component's velocities are large due to the faintness and
large rotational broadening of the component.

\subsubsection{LP~734-34AB}
\label{subsubsec:lp734}

\begin{figure*}
\minipage{0.50\textwidth}
\includegraphics[scale=.47,angle=0]{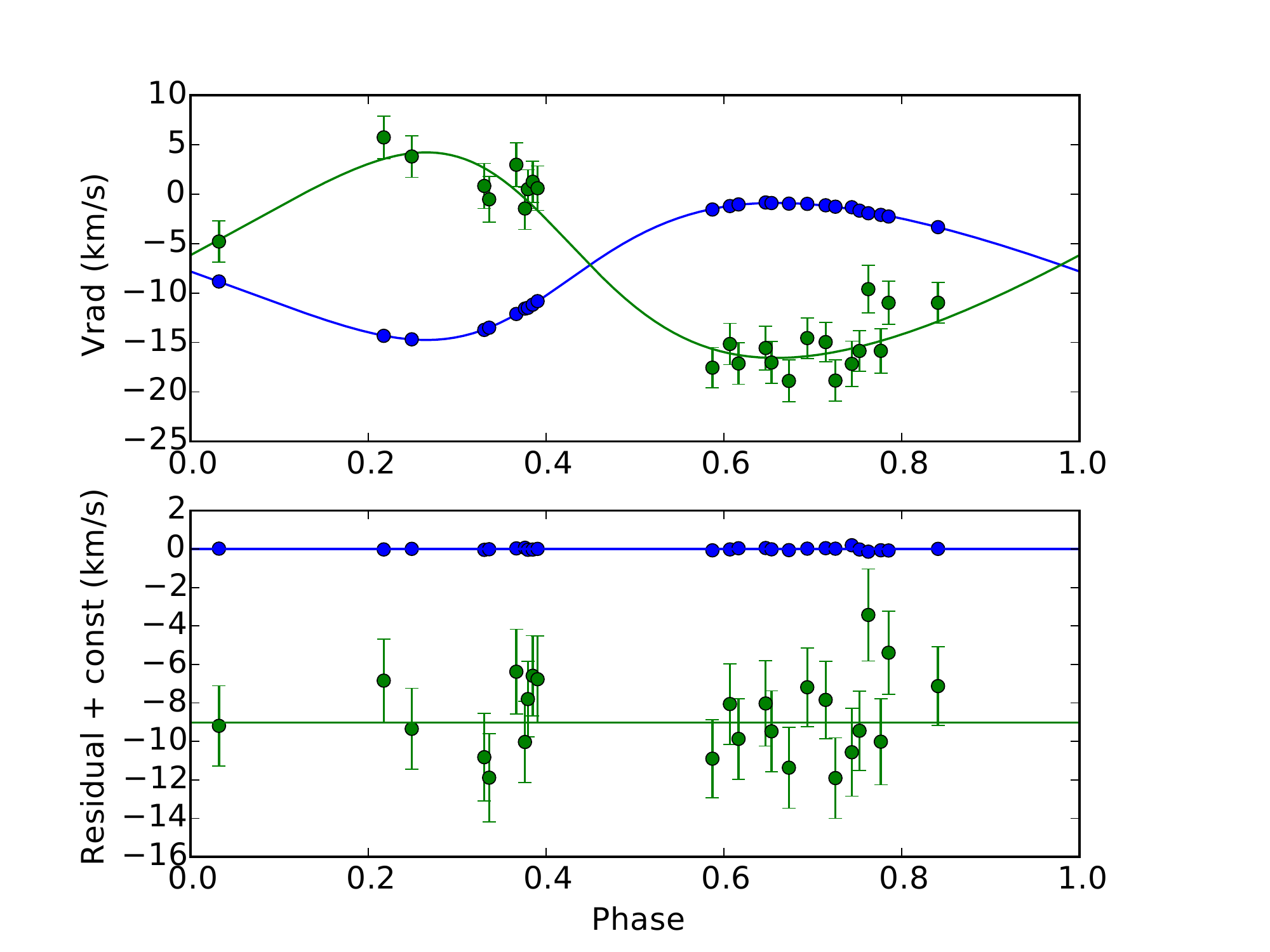}
\endminipage\hfill
\minipage{0.50\textwidth}
\includegraphics[scale=.47,angle=0]{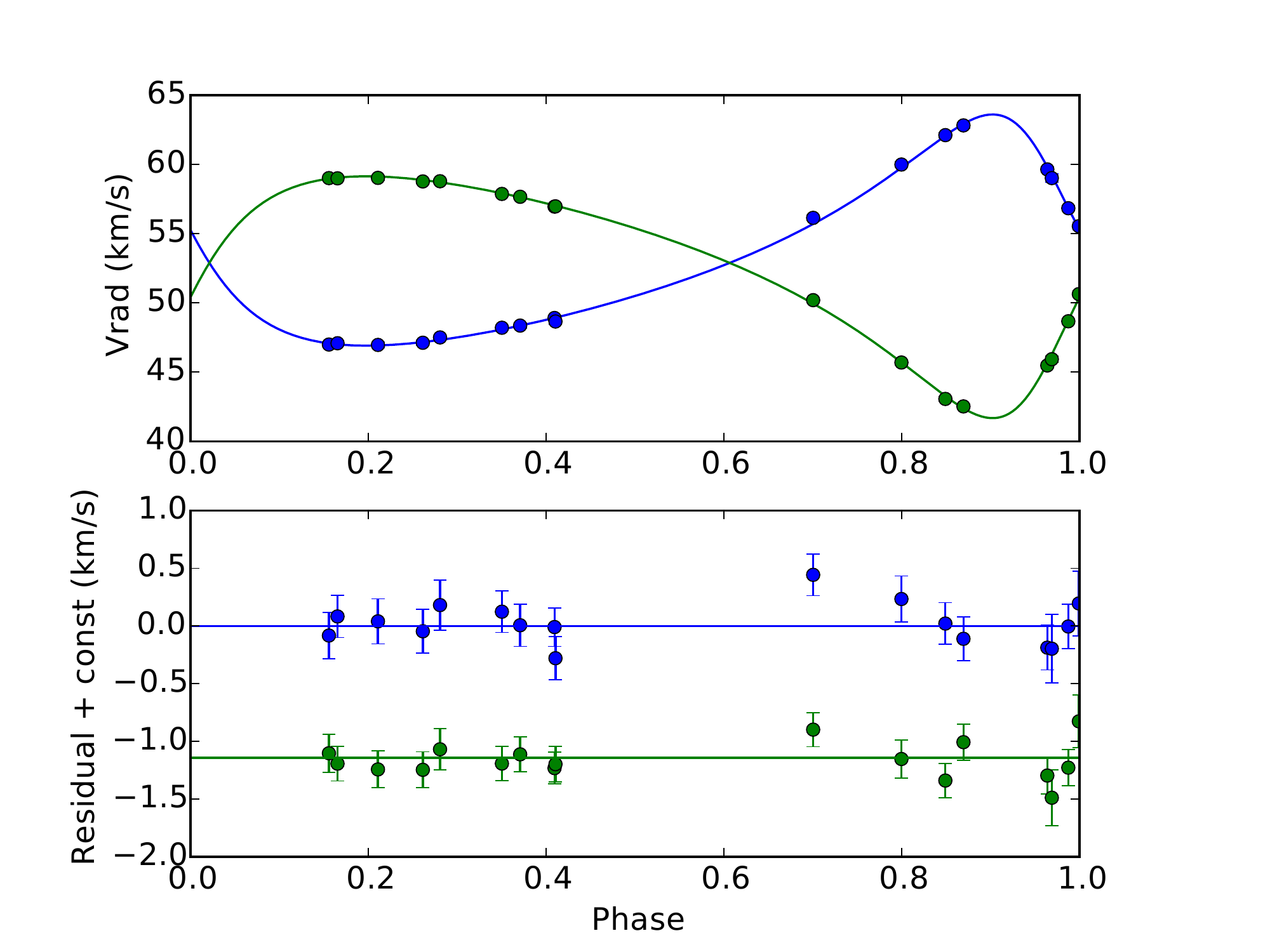}
\endminipage\hfill
\caption{The orbits of 2MA~0930+0227AB (left) and LP~734-34AB (right). \label{fig:2ma0930_lp734_orbit}}
\end{figure*}

This new double-lined spectroscopic binary had no existing spectrum or
parallax in the literature when it was added to our target list. We
acquired 24 spectra of LP~734-34AB with a S/N per spectral resolution
element of 8-29. We estimate a light ratio of 0.905 to derive the
velocities with negligible broadening. The removal of six epochs with
poorly-separated velocities, as well as the flipping of the
components' velocities for the ninth, eleventh, and sixteenth
observations, yields a sensible fit for the orbit, which we show in
the right panel of Figure \ref{fig:2ma0930_lp734_orbit}. We find an
orbital period of $33.6551 \pm 0.0046$ days, an eccentricity of $0.423
\pm 0.010$, and a mass ratio of $0.956 \pm 0.010$ for this system.

\subsubsection{GJ~1182AB}
\label{subsubsec:gj1182}

This object was noted in \citet{Jenkins(2009)} as having a possible
binary component, and an orbit for this system was reported in
\citet{Baroch(2018)}. As with GJ~1029, we independently discovered
this double-lined system in 2017 and began measuring the orbit.  We
obtained 17 spectra of GJ~1182AB with S/N per spectral resolution
element of 11-24 from which we derive a light ratio of 0.197 with no
correction for rotational broadening applied. We discarded three
epochs with insufficient velocity separation, and measure a $154.23
\pm 0.51$-day orbital period, in agreement with that reported in
\citet{Baroch(2018)}. We find an eccentricity of $0.5362 \pm 0.0022$
and a mass ratio of $0.6606 \pm 0.0095$ for this system. We show the
orbit in the left panel of Figure \ref{fig:gj1182_g258_orbit}.

\subsubsection{G~258-17AB}
\label{subsubsec:g258}

\begin{figure*}
\minipage{0.50\textwidth}
\includegraphics[scale=.47,angle=0]{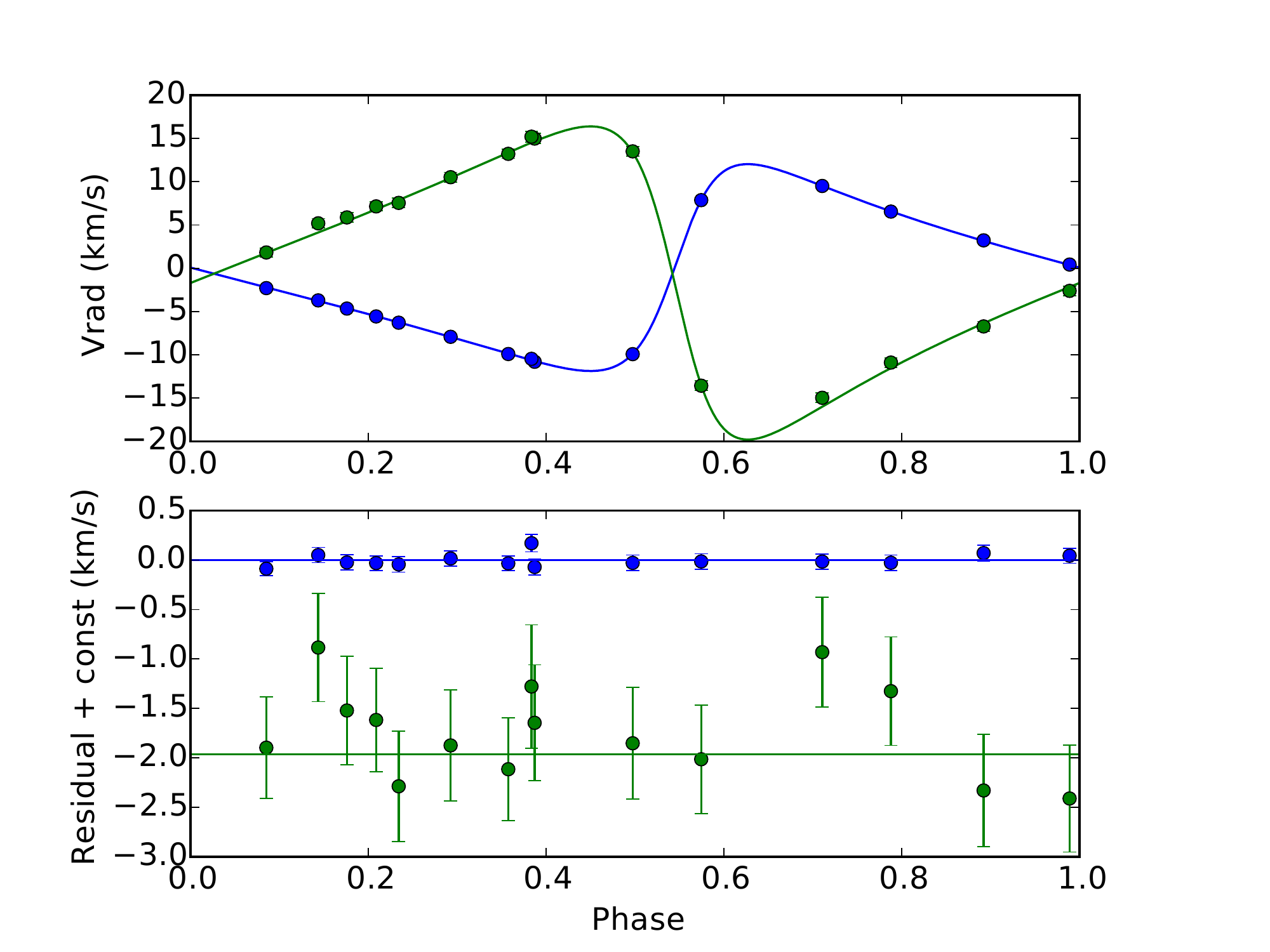}
\endminipage\hfill
\minipage{0.50\textwidth}
\includegraphics[scale=.47,angle=0]{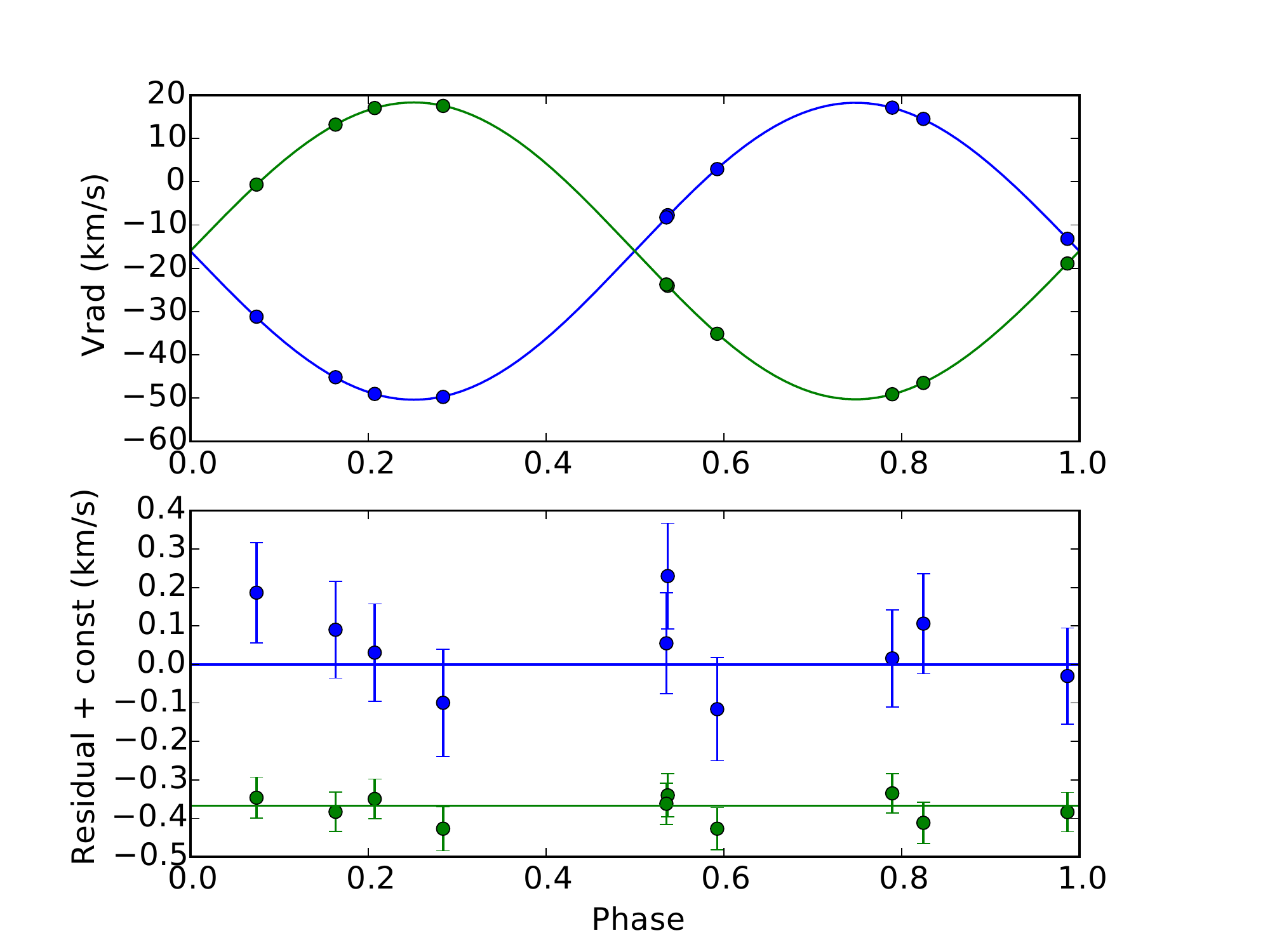}
\endminipage\hfill
\caption{The orbital fits of GJ~1182AB (left) and G~258-17AB (right). \label{fig:gj1182_g258_orbit}}
\end{figure*}

G~258-17 is a wide companion to the more massive star HD~161897 and is
found at an angular separation of 89\farcs8 at a position angle of
28.3\arcdeg ~from the primary, as noted by \citet{Tokovinin(2014)}.
We initially overlooked an existing $Hipparcos$ parallax
\citep{vanLeeuwen(2007)} for HD~161897 of $33.30\pm0.47$ mas and
included the star in our sample based on the parallax by
\citet{Dittmann(2014)}; however, the {\it Gaia} DR2 parallax for
G~258-17 of $33.5254\pm0.0492$ mas is in agreement with that by
\citet{vanLeeuwen(2007)}. We report the discovery of a near-equal
luminosity companion to G~258-17, making the system a hierarchical
triple. We acquired 11 spectra of G~258-17AB with a S/N per spectral
resolution element of 19-32. We see negligible rotational broadening
in the spectra and use a flux ratio of 0.986 to calculate the
velocities. We omitted one epoch due to insufficient velocity
separation between the components. Because {\sc todcor} can sometimes
confuse components in nearly-equal-luminosity systems, we reversed the
velocities of the components for the first, third, and fourth
observations, assuming that the brightest component was the
primary. We find an orbital period of $4.741475 \pm 0.000018$ days, an
eccentricity of $0.00495 \pm 0.00096$, and a mass ratio of $1.0003 \pm
0.0019$. We show our orbital fit to the velocities in the right panel
of Figure \ref{fig:gj1182_g258_orbit}.

\subsubsection{LTT~7077AB}
\label{subsubsec:ltt07077}

\begin{figure*}
\minipage{0.50\textwidth}
\includegraphics[scale=.47,angle=0]{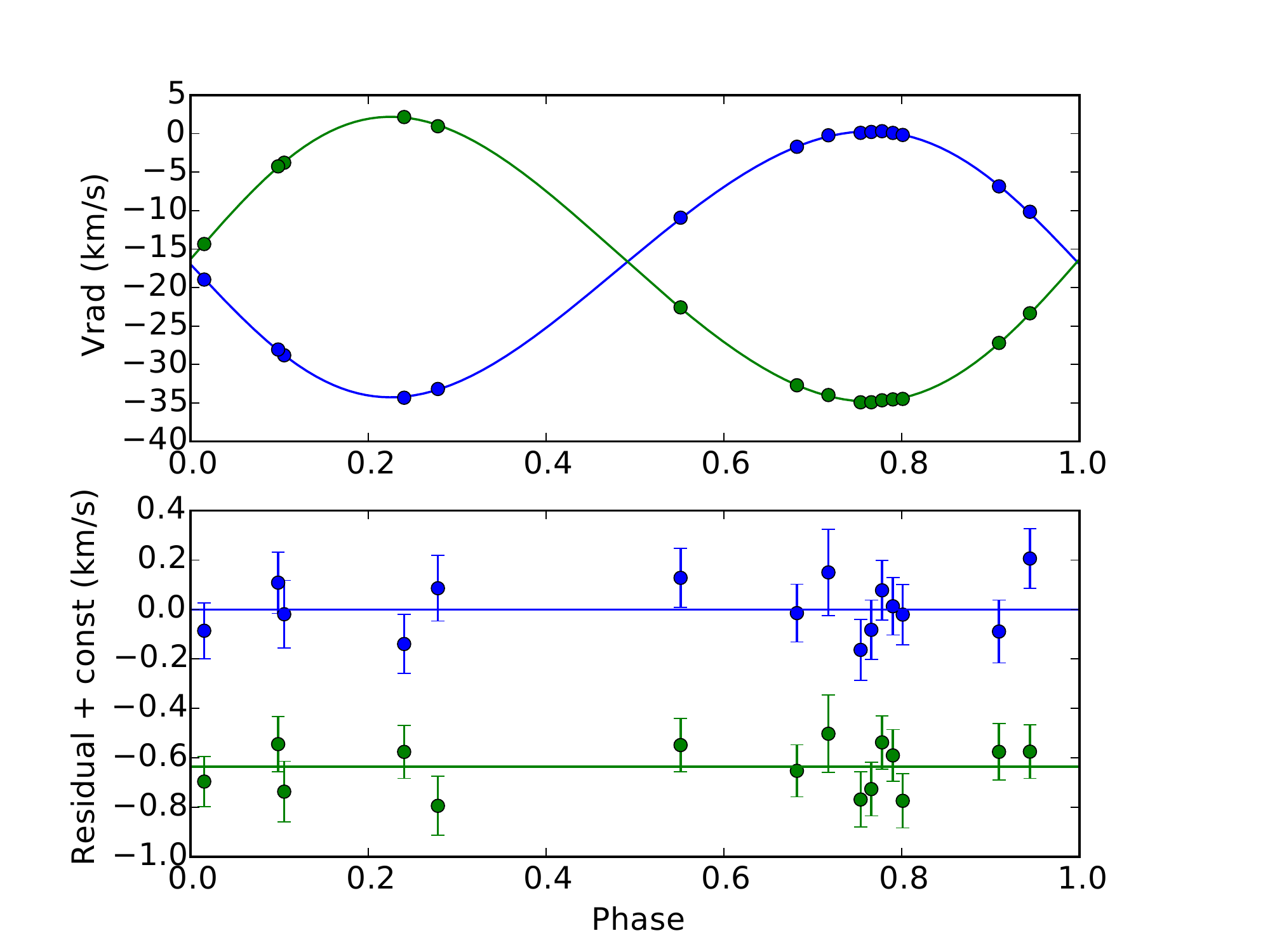}
\endminipage\hfill
\minipage{0.50\textwidth}
\includegraphics[scale=.47,angle=0]{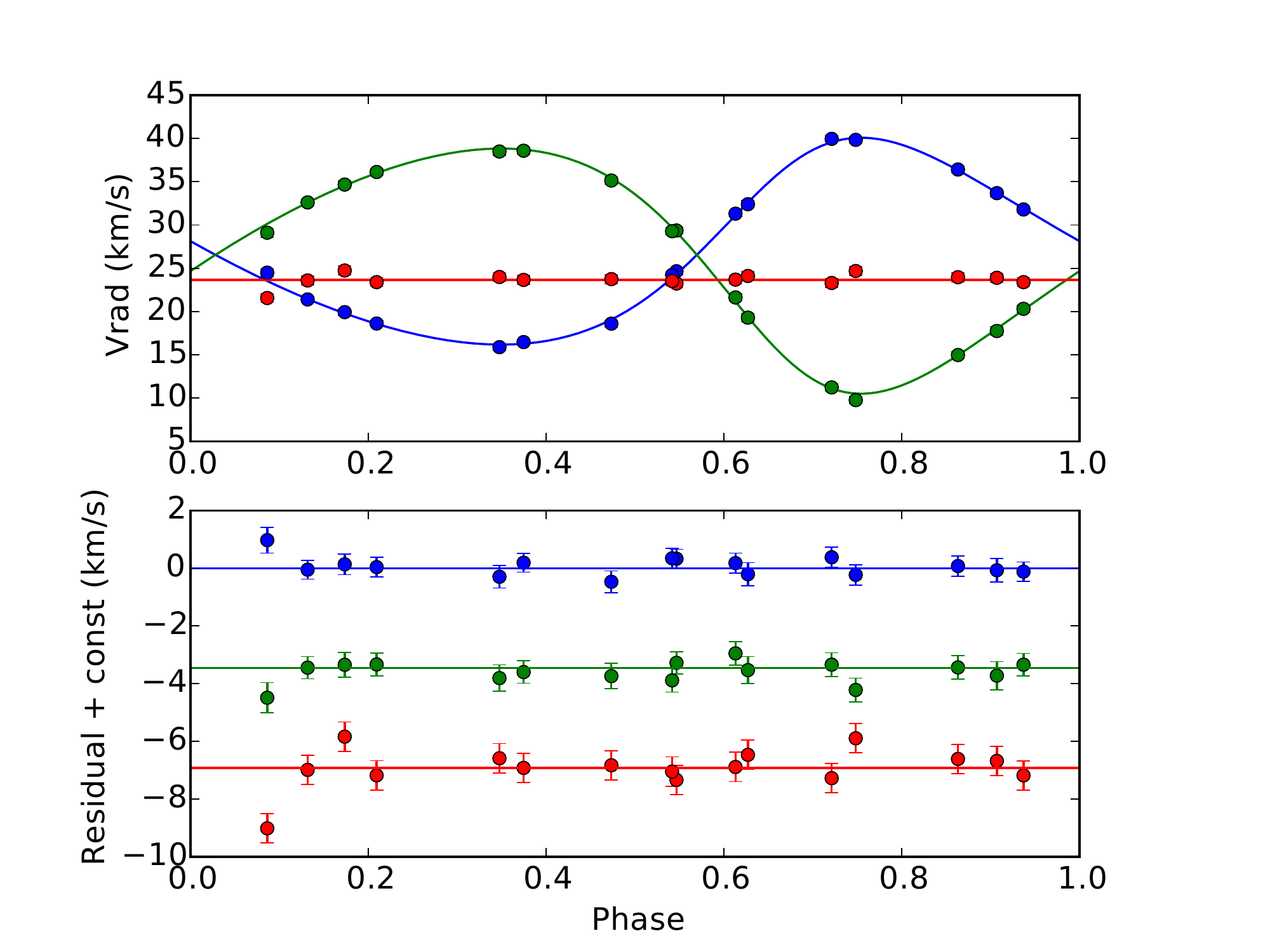}
\endminipage\hfill
\caption{The orbital fits of LTT~7077AB (left) and LP~655-43ABC (right). \label{fig:ltt07077_lp655_orbit}}
\end{figure*}

This object was not initially included in our sample because, although
it is nearer than 15 pc, the estimated mass of the primary from $M_K$
is 0.35 \mdot. However, this system was noted by \citet{Malo(2014)} to
be a double-lined spectroscopic binary, so we chose to observe this
system under the assumption that a deblended $K$ magnitude and/or a
measured orbit would result in a primary mass within our range of
interest. We obtained 17 spectra of LTT~7077AB with S/N per spectral
resolution element of 6-21. We assumed zero rotational broadening and
find a light ratio of 0.901. We discarded two data points because the
velocities were not widely separated and reversed the components'
velocities for the first observation.  We measure an orbital period of
$83.926 \pm 0.032$ days, an eccentricity of $0.0640 \pm 0.0021$, and a
mass ratio of $0.9341 \pm 0.0031$ for LTT~7077AB. We show the orbit in
the left panel of Figure \ref{fig:ltt07077_lp655_orbit}.

\subsection{Triple-lined System: LP~655-43ABC}
\label{subsec:triple}

This is a new triple-lined system. We gathered 16 spectra of
LP~655-43ABC with S/N per spectral resolution element of 13-26. The
three components are resolved for only three epochs. We derive a light
ratio of 0.76 for the secondary-to-primary pair of components and 0.40
for the tertiary-to-primary pair, which we used to calculate the
velocities.  We fix the velocities for the third component to be
between 22 and 26 km s$^{-1}$. We see negligible rotational broadening
for any of the components. We measure for LP~655-43AB an orbital
period of $18.3715 \pm 0.0084$ days, an eccentricity of $0.1994 \pm
0.0078$, and a mass ratio of $0.844 \pm 0.014$. We show the orbit in
Figure \ref{fig:ltt07077_lp655_orbit}. The velocity uncertainties for
the C component shown in Figure \ref{fig:ltt07077_lp655_orbit} are the
standard deviation of the calculated velocities, as we did not fit
this component in our MCMC analysis. A photometric rotation period
reported by \citet{Newton(2018)} of 18.38 days, in agreement with the
orbital period of 18.38 days for the inner SB2, indicates that the
orbit of the inner pair has synchronized.

Because the velocities of the C component do not significantly vary,
we also compared the POSS-1 red plate image taken 1953.933 to the UK
Schmidt $I-$band image taken 2001.804 to ensure that the SB2 had not
moved on top of a background star. No other star is seen at the
current position of the system. In addition, there is only one Gaia
DR2 data point at the location of the system, so C is not a background
star with a separate (and discrepant) parallax.

\startlongtable
\begin{deluxetable*}{llclc}
\tabletypesize{\small}
\tablewidth{0pt}
\tablecaption{Orbital Elements for Binaries \label{tab:orb_el}}
\tablecolumns{5}
\tablenum{5}

\tablehead{
\colhead{Name}            & 
\colhead{MCMC Parameter}    &
\colhead{MCMC Value}       &
\colhead{Derived Parameter}  &
\colhead{Derived Value}
}
\startdata
\hline
GJ~1029AB    &                                      &                              &                                       &                        \\ 
\hline                                                                
             & $e \cos \omega$                      & $-0.3280 \pm 0.0069$        &  $e$                                    & $0.3786 \pm 0.0067$     \\      
             & $e \sin \omega$                      &  $-0.1891 \pm 0.0084$       &  $\omega$ (deg)                         & $210.0 \pm 1.3$        \\      
             & $T_0$ (BJD)                          & $2457741.08 \pm 0.56$       &  $a \sin i$ (AU)                        & $0.1331 \pm 0.0027$    \\    
             & $P$ (days)                           & $95.76 \pm 0.18$            &  $(M_1+M_2) \sin^3 i$ (${\rm M}_\odot$) & $0.03430 \pm 0.00210$  \\    
             & $q$                                  & $0.722 \pm 0.026$           &  $M_1 \sin^3 i$ (${\rm M}_\odot$)       & $0.01991 \pm 0.00150$  \\
             & $\gamma$ (km s$^{-1}$)               & $-11.189 \pm 0.036$         &  $M_2 \sin^3 i$ (${\rm M}_\odot$)       & $0.01439 \pm 0.00061$  \\  
             & $(K_1+K_2)$ (km s$^{-1}$)            & $16.34 \pm 0.34$            &  $T_{\rm peri}$ (BJD)                   & $2457761.58 \pm 0.54$  \\ 
             & $\sigma_1$ (km s$^{-1}$)             & $0.099 \pm 0.028$           &  \nodata                                &  \nodata               \\   
             & $\sigma_2$ (km s$^{-1}$)             & $0.774 \pm 0.161$           &  ${\alpha}_{ast}$ (mas)                 &  $4.6$              \\
\hline
LP~655-43AB &                                      &                              &                                       &                        \\
\hline                                                             
             & $e \cos \omega$                      & $0.1324 \pm 0.0082$         &  $e$                                    & $0.1994 \pm 0.0078$   \\      
             & $e \sin \omega$                      & $-0.1487 \pm 0.0088$        &  $\omega$ (deg)                         & $311.7 \pm 2.7$       \\  
             & $T_0$ (BJD)                          & $2458036.080 \pm 0.069$     &  $a \sin i$ (AU)                        & $0.04322 \pm 0.00036$   \\ 
             & $P$ (days)                           & $18.3715 \pm 0.0084$        &  $(M_1+M_2) \sin^3 i$ (${\rm M}_\odot$) & $0.03191 \pm 0.00080$    \\   
             & $q$                                  & $0.844 \pm 0.014$           &  $M_1 \sin^3 i$ (${\rm M}_\odot$)       & $0.01731 \pm 0.00046$  \\     
             & $\gamma$ (km s$^{-1}$)                & $26.550 \pm 0.072$         &  $M_2 \sin^3 i$ (${\rm M}_\odot$)       & $0.01460 \pm 0.00038$  \\  
             & $(K_1+K_2)$ (km s$^{-1}$)             & $26.12 \pm 0.22$           &  $T_{\rm peri}$ (BJD)                   & $2458029.89 \pm 0.12$  \\
             & $\sigma_1$ (km s$^{-1}$)             & $0.288 \pm 0.060$           &  \nodata                                & \nodata                  \\  
             & $\sigma_2$ (km s$^{-1}$)             & $0.339 \pm 0.071$           &  ${\alpha}_{ast}$ (mas)                 & 0.2              \\  
\hline
LTT~11586Ac  &                                      &                              &                                       &                        \\      
\hline                                                                
             & $e \cos \omega$                      & $0.1202 \pm 0.0090$         & $e$                                   & $0.5029 \pm 0.0055$     \\    
             & $e \sin \omega$                      & $-0.4882 \pm 0.0074$        & $\omega$ (deg)                        & $283.8 \pm 1.2$       \\    
             & $T_0$ (BJD)                          & $2458108.670 \pm 0.073$     & $T_{\rm peri}$ (BJD)                  & $2458102.629 \pm 0.039$  \\ 
             & $P$ (days)                           & $15.04547 \pm 0.00041$      & $a_1 \sin i$ (AU)                     & $0.010529 \pm 0.000095$   \\  
             & $\gamma$ (km s$^{-1}$)               & $21.112 \pm 0.024$          & $f_1(M)$ (${\rm M}_\odot$)          & $0.000688 \pm 0.000019$  \\ 
             & $\dot{\gamma}$ (km s$^{-1}$ day$^{-1}$)   & $-0.002323 \pm 0.000093$  & $q_{\rm min}$                      & $0.1462 \pm 0.0014$      \\
             & $\ddot{\gamma}$ (km s$^{-1}$ day$^{-1}$ day$^{-1}$)  & $0.00000172 \pm 0.00000035$ &  $a_{\rm min}$ (AU)   & $0.082527 \pm 0.000035$ \\    
             & $K$ (km s$^{-1}$)                    & $8.81 \pm 0.11$             &  $M_{2,{\rm min}}$ (${\rm M}_\odot$)   & $0.04226 \pm 0.00042$     \\  
             & $\sigma$ (km s$^{-1}$)               & $0.058 \pm 0.014$           & $M_{2,{\rm min}}$ (${\rm M}_{\rm Jup}$)  &  $44.27 \pm 0.44$     \\     
             & \nodata                              & \nodata                     & ${\alpha}_{ast}$ (mas)                 &  $> 0.1$              \\  
\hline
LHS~1817Ab     &                                      &                              &                                       &                        \\      
\hline                                                                
             & $e \cos \omega$                      & $-0.0002 \pm 0.0033$           &  $e$                                     & $0.0063 \pm 0.0031$            \\
             & $e \sin \omega$                      & $0.0053 \pm 0.0032$            &  $\omega$ (deg)                          & $93 \pm 37$                  \\
             & $T_0$ (BJD)                          &  $2458357.72859 \pm 0.00029$   &  $T_{\rm peri}$ (BJD)                    & $2458357.731 \pm 0.032$  \\ 
             & $P$ (days)                           & $0.30992678 \pm 0.00000048$    &  $a_1 \sin i$ (AU)                       & $0.0035702 \pm 0.0000089$         \\
             & $\gamma$ (km s$^{-1}$)               & $-0.71 \pm 0.35$               &  $f_1(M)$ (${\rm M}_\odot$)              & $0.06321 \pm 0.00048$   \\  
             & $K$ (km s$^{-1}$)                    & $125.33 \pm 0.31$              &  $q_{\rm min}$                           & $0.9343 \pm 0.0035$    \\  
             & $\sigma$ (km $^{-1}$)                & $0.75 \pm 0.11$                &  $a_{\rm min}$ (AU)                      & $0.0073917 \pm 0.0000044$\\  
             & \nodata                              & \nodata                        &  $M_{2,{\rm min}}$ (${\rm M}_\odot$)     & $0.2709 \pm 0.0010$ \\  
             & \nodata                              & \nodata                        &  $M_{2,{\rm min}}$ (${\rm M}_{\rm Jup}$) &  $283.8 \pm 1.0$ \\  
             & \nodata                              & \nodata                        & ${\alpha}_{ast}$ (mas)                   &  $> 0.1$              \\  
\hline
2MA~0930$+$0227AB                                  &                               &                                       &                        \\
\hline
             & $e \cos \omega$                      & $-0.0950 \pm 0.0048$         & $e$                                  & $0.1928 \pm 0.0046$   \\
             & $e \sin \omega$                      & $-0.1677 \pm 0.0037$         & $\omega$ (deg)                       & $240.5 \pm 1.2$  \\
             & $T_0$ (BJD)                          & $2457910.7 \pm 2.6$          & $a \sin i$ (AU)                      & $1.433 \pm 0.045$    \\
             & $P$ (days)                           & $916.8 \pm 2.5$              & $(M_1+M_2) \sin^3 i$ (${\rm M}_\odot$) & $0.467 \pm 0.044$  \\
             & $q$                                  & $0.667 \pm 0.035$            & $M_1 \sin^3 i$ (${\rm M}_\odot$)     & $0.280 \pm 0.032$    \\
             & $\gamma$ (km s$^{-1}$)               & $-7.161 \pm 0.025$           & $M_2 \sin^3 i$ (${\rm M}_\odot$)     & $0.187 \pm 0.012$    \\
             & $(K_1+K_2)$ (km s$^{-1}$)            & $17.32 \pm 0.54$             & $T_{\rm peri}$ (BJD)                 & $2458262.3 \pm 2.6$ \\
             & $\sigma_1$ (km s$^{-1}$)             & $0.0580 \pm 0.0094$          & \nodata                              & \nodata                    \\
             & $\sigma_2$ (km s$^{-1}$)             & $1.8149 \pm 0.2633$          & ${\alpha}_{ast}$ (mas)                 & $7$                         \\
\hline
LP~734-34AB                                         &                              &                                       &                       \\   
\hline                                                                
             & $e \cos \omega$                      & $0.2839 \pm 0.0092$          &   $e$                                  & $0.423 \pm 0.010$   \\  
             & $e \sin \omega$                      & $0.3139 \pm 0.0080$          &   $\omega$ (deg)                       & $47.89 \pm 0.93$      \\  
             & $T_0$ (BJD)                          & $2457868.220 \pm 0.066$      &   $a \sin i$ (AU)                      & $0.04786 \pm 0.00034$ \\  
             & $P$ (days)                           & $33.6551 \pm 0.0046$         &   $(M_1+M_2) \sin^3 i$ (${\rm M}_\odot$) & $0.01292 \pm 0.00028$ \\
             & $q$                                  & $0.956 \pm 0.010$            &   $M_1 \sin^3 i$ (${\rm M}_\odot$)     & $0.00661 \pm 0.00014$ \\
             & $\gamma$ (km s$^{-1}$)               & $52.890 \pm 0.031$           &   $M_2 \sin^3 i$ (${\rm M}_\odot$)     & $0.00631 \pm 0.00014$  \\ 
             & $(K_1+K_2)$ (km s$^{-1}$)            & $17.08 \pm 0.19$             &   $T_{\rm peri}$ (BJD)                 & $2457866.693 \pm 0.061$\\ 
             & $\sigma_1$ (km s$^{-1}$)             & $0.156 \pm 0.036$            &    \nodata                             & \nodata                 \\
             & $\sigma_2$ (km s$^{-1}$)             & $0.128 \pm 0.031$            & ${\alpha}_{ast}$ (mas)                   & $0.1$               \\
\hline                                                              
G~123-45Ab  &                                      &                              &                                        &                          \\
\hline                                                               
             & $e \cos \omega$                      & $-0.1153 \pm 0.0022$         &  $e$                                   & $0.3758 \pm 0.0024$     \\
             & $e \sin \omega$                      & $0.3577 \pm 0.0025$          &  $\omega$ (deg)                        & $107.86 \pm 0.35$       \\    
             & $T_0$ (BJD)                          & $2457754.486 \pm 0.063$      &  $T_{\rm peri}$ (BJD)                  & $2457755.218 \pm 0.069$ \\ 
             & $P$ (days)                           & $34.7557 \pm 0.0041$         &  $a_1 \sin i$ (AU)                     & $0.021874 \pm 0.000046$ \\
             & $\gamma$ (km s$^{-1}$)               & $7.431 \pm 0.012$            &  $f_1(M)$ (${\rm M}_\odot$)            & $0.0011559 \pm 0.0000073$ \\
             & $K$ (km s$^{-1}$)                    & $7.388 \pm 0.017$            &  $q_{\rm min}$                         & $0.20903 \pm 0.00049$      \\
             & $\sigma$ (km s$^{-1}$)               & $0.0382 \pm 0.0075$          &  $a_{\rm min}$ (AU)                    & $0.126519 \pm 0.000020$   \\
             & \nodata                              & \nodata                      &  $M_{2,{\rm min}}$ (${\rm M}_\odot$)   & $0.038671 \pm 0.000091$  \\
             & \nodata                              & \nodata                      &  $M_{2,{\rm min}}$ (${\rm M}_{\rm Jup}$)  & $40.510 \pm 0.096$ \\      
             & \nodata                              & \nodata                        & ${\alpha}_{ast}$ (mas)                 &  $> 0.3$         \\  
\hline       
GJ~1182AB    &                                      &                              &                                        &                           \\
\hline                                                              
             & $e \cos \omega$                      & $0.0593 \pm 0.0055$          & $e$                                    & $0.5362 \pm 0.0022$ \\        
             & $e \sin \omega$                      & $-0.5329 \pm 0.0021$         & $\omega$ (deg)                         & $276.35 \pm 0.58$ \\           
             & $T_0$ (BJD)                          & $2457782.9 \pm 1.6$          & $a \sin i$ (AU)                        & $0.3596 \pm 0.0040$ \\        
             & $P$ (days)                           & $154.23 \pm 0.51$            & $(M_1+M_2) \sin^3 i$ (${\rm M}_\odot$) & $0.2608 \pm 0.0077$ \\      
             & $q$                                  & $0.6606 \pm 0.0095$          & $M_1 \sin^3 i$ (${\rm M}_\odot$)       & $0.1570 \pm 0.0054$ \\      
             & $\gamma$ (km $^{-1}$)                & $-0.625 \pm 0.041$           & $M_2 \sin^3 i$ (${\rm M}_\odot$)       & $0.1038 \pm 0.0023$ \\
             & $(K_1+K_2)$ (km $^{-1}$)             & $30.05 \pm 0.29$             & $T_{\rm peri}$ (BJD)                   & $2457713.3 \pm 1.5$ \\
             & $\sigma_1$ (km $^{-1}$)              & $0.068 \pm 0.017$            & \nodata                                & \nodata             \\
             & $\sigma_2$ (km $^{-1}$)              & $0.484 \pm 0.099$            & ${\alpha}_{ast}$ (mas)                 & $5.8$            \\
\hline
G~258-17AB   &                                      &                              &                                        &                           \\
\hline                                                              
             & $e \cos \omega$                      & $-0.00118 \pm 0.00070$       & $e$                                  & $0.00495 \pm 0.00096$        \\   
             & $e \sin \omega$                      & $-0.00475 \pm 0.00106$       & $\omega$ (deg)                       & $256.0 \pm 9.5$               \\   
             & $T_0$ (BJD)                          & $2457826.5789 \pm 0.0017$    & $a \sin i$ (AU)                      & $0.029903 \pm 0.000034$     \\  
             & $P$ (days)                           & $4.741475 \pm 0.000018$      & $(M_1+M_2) \sin^3 i$ (${\rm M}_\odot$) & $0.15866 \pm 0.00053$      \\
             & $q$                                  & $1.0003 \pm 0.0019$          & $M_1 \sin^3 i$ (${\rm M}_\odot$)     & $0.07932 \pm 0.00023$         \\
             & $\gamma$ (km $^{-1}$)                & $-16.019 \pm 0.018$          & $M_2 \sin^3 i$ (${\rm M}_\odot$)     & $0.07934 \pm 0.00032$         \\
             & $(K_1+K_2)$ (km $^{-1}$)             & $68.608 \pm 0.077$           & $T_{\rm peri}$ (BJD)                 & $2457828.76 \pm 0.13$       \\ 
             & $\sigma_1$ (km $^{-1}$)              & $0.117 \pm 0.031$            & \nodata                              & \nodata                     \\
             & $\sigma_2$ (km $^{-1}$)              & $0.048 \pm 0.017$            & ${\alpha}_{ast}$ (mas)                 & \nodata                    \\
\hline
LTT~7077AB   &                                      &                              &                                      &                           \\
\hline
             & $e \cos \omega$                      & $-0.0186 \pm 0.0019$         & $e$                                  & $0.0640 \pm 0.0021$ \\
             & $e \sin \omega$                      & $0.0612 \pm 0.0021$          & $\omega$ (deg)                       & $106.9 \pm 1.6$     \\
             & $T_0$ (BJD)                          & $2458290.87 \pm 0.12$        & $a \sin i$ (AU)                      & $0.27559 \pm 0.00048$ \\
             & $P$ (days)                           & $83.926 \pm 0.032$           & $(M_1+M_2) \sin^3 i$ (${\rm M}_\odot$) & $0.3964 \pm 0.0020$ \\
             & $q$                                  & $0.9341 \pm 0.0031$          & $M_1 \sin^3 i$ (${\rm M}_\odot$)     & $0.2049 \pm 0.0011$   \\
             & $\gamma$ (km s$^{-1}$)               & $-16.661 \pm 0.023$          & $M_2 \sin^3 i$ (${\rm M}_\odot$)     & $0.1915 \pm 0.0011$   \\
             & $(K_1+K_2)$ (km s$^{-1}$)            & $35.795 \pm 0.062$           & $T_{\rm peri}$ (BJD)                 & $2458294.34 \pm 0.38$ \\
             & $\sigma_1$ (km s$^{-1}$)             & $0.102 \pm 0.021$            &  \nodata                             & \nodata                \\
             & $\sigma_2$ (km s$^{-1}$)             & $0.092 \pm 0.020$            & ${\alpha}_{ast}$ (mas)                 & $0.5$                \\
\enddata  
\tablecomments{The uncertainty on the systemic velocity $\gamma$ for each system does not include the systematic uncertainty of 0.5 km s$^{-1}$ from the Barnard's
  Star template radial velocity, which should be added in quadrature
  when calculating the total uncertainty on $\gamma$.}
\end{deluxetable*}

\section{Discussion} \label{sec:discussion}

We have measured the spectroscopic orbits of eleven binaries with
mid-to-late M dwarf components, including the well-known SB2
GJ~268AB. Within 15 pc, we contribute to the currently known M dwarf
multiples population orbital parameters for two new possible brown
dwarf and one new M dwarf companion in three systems (G~123-45,
LTT~11586, LTT~7077). At distances 15-25 pc, we add system parameters
for a new white dwarf and two new M dwarf components in three systems
(LHS~1817, 2MA~0930, LP~734-34). And beyond 25 pc, we contribute
orbital parameters for two new M dwarf companions in two systems
(G~258-17, LP~655-43). In addition, we present RVs for the components
of L~870-44 and for the B and C components of LTT~11586 and LP~655-43,
respectively.

If the three possible sub-stellar components that we have discovered
to-date (including LHS~1610b, which we reported in
\citet{Winters(2018)}) are found to be indeed sub-stellar-mass
objects, this would represent a doubling from 0.8\% (3/376) to 1.6\%
(6/376) in the number of mid-to-late M dwarf primaries known to host
brown dwarf companions within 15 pc. It is likely that they eluded
detection because previous radial velocity surveys of these types of
stars have typically obtained only a single observation.

Our discovery of a new, nearby, white dwarf - M dwarf system
(LHS~1817Ab) provides an intriguing glimpse into stellar
evolution. The M dwarf is unlikely to have been able to survive the
violent environment that resulted in the creation of the white dwarf
at the current separation ($a_{\rm min} = 0.00766$ AU). Thus, the
system evolved to the observed configuration over time. This system
joins the rare examples of nearby M dwarf - white dwarf systems, a
combination found to-date in only 2\% of the known multiple systems
with M dwarf primaries within 25 pc \citep{Winters(2019a)}. The white
dwarf component is a new member of the white dwarf population within
25 pc \citep{Holberg(2016),Subasavage(2017),Hollands(2018)}.

Some of the new systems presented here will be benchmark systems
because of the possibility of deriving the true masses for the
components. While recent results from the Robo-AO survey indicate that
the {\it Gaia} DR2 does not resolve binaries with separations less
than roughly 1\arcsec ~\citep{Ziegler(2018)}, the final {\it Gaia}
data release will publish astrometric orbits for binary systems. We
therefore estimate in the $R_{\sc KC}$ filter the magnitudes of the
astrometric perturbations ${\alpha}_{ast}$ of the systems reported
here. For double-lined systems, we begin with an estimated $\Delta K$
magnitude between the components. We then varied the magnitude
difference and calculated the component masses using the MLR by
\citet{Benedict(2016)} and the resulting mass ratio until it agreed
with the mass ratio from our orbital solution. We then converted the
final $\Delta K$ to $\Delta R_{\sc KC}$ using the relations in
\citet{Riedel(2014)} and followed the prescription in
\citet{vandeKamp(1975)} to calculate the magnitude of the astrometric
perturbation. For the single-lined systems, we use $q_{min}$, $a_{1}
\sin i$ converted to arcseconds via the parallax, and assume a $\Delta
K$ of 10 magnitudes. Thus, our estimates for the single-lined systems
are lower limits. We estimate the magnitudes of astrometric
perturbations for our systems to range between 0.0 mas to 7.0 mas and
list them in Table \ref{tab:orb_el}. The astrometric perturbations of
the near-equal-mass systems with small separations between components
(LP~734-34 and G~258-17) will likely not be detected by {\it Gaia}, as
the shift in the position of the photocenter is below the anticipated
astrometric precision for binaries with small magnitude differences
between the components \citep[0.2 mas;][]{Lindegren(2018)}. GJ~1029,
GJ~268, 2MA~0930, GJ~1182, and LTT~7077 have estimated perturbations
of 4.6, 0.7, 7.0, 5.8, and 0.5 mas, respectively. These will be
benchmark systems, as the astrometric orbital solutions from {\it
  Gaia} will provide the orbital inclinations that will enable the
calculation of the true masses of each component.

The orbits that we have measured help to fill in some of the gaps in
our knowledge of the period, separation, eccentricity, and mass ratio
distributions for M dwarf multiple stars. \citet{Duchene(2013)} find
that the separation distribution for M dwarfs across all spectral
sub-types peaks at 5.3 AU, whereas recent results from
\citet{Winters(2019a)} found peaks of 4 AU and 20 AU for the
volume-limited 10 pc and 25 pc M dwarf samples, respectively. The true
answer is likely closer to 5 AU, but the closest M dwarfs have not yet
been comprehensively surveyed with the multi-epoch, high-resolution
techniques necessary to detect companions at such separations. We are
resolving this incompleteness with our high-resolution spectroscopic
and speckle imaging surveys (Winters et al., in prep). A full analysis of
orbital parameter distributions for mid-to-late M dwarfs is beyond the
scope of this paper, but will be addressed once the southern
hemisphere portion of our spectroscopic survey is complete.

\acknowledgments

The authors thank the anonymous referee for their prompt response and
for their comments and suggestions which improved the manuscript. We
thank Matthew Payne, Samuel Quinn, Guillermo Torres, Warren Brown,
John Subasavage, and Emily Leiner for illuminating discussions.

The MEarth Team gratefully acknowledges funding from the David and
Lucille Packard Fellowship for Science and Engineering (awarded to
D.C.). This material is based upon work supported by the National
Science Foundation under grants AST-0807690, AST-1109468, AST-1004488
(Alan T. Waterman Award), and AST-1616624. This work is made possible
by a grant from the John Templeton Foundation. The opinions expressed
in this publication are those of the authors and do not necessarily
reflect the views of the John Templeton Foundation. This material is
based upon work supported by the National Aeronautics and Space
Administration under Grant No. 80NSSC18K0476 issued through the XRP
Program. AAM is supported by NSF Graduate Research Fellowship grant
DGE1745303.

This work has made use of data from the European Space Agency (ESA)
mission {\it Gaia} (\url{https://www.cosmos.esa.int/gaia}), processed
by the {\it Gaia} Data Processing and Analysis Consortium (DPAC,
\url{https://www.cosmos.esa.int/web/gaia/dpac/consortium}). Funding
for the DPAC has been provided by national institutions, in particular
the institutions participating in the {\it Gaia} Multilateral
Agreement. This work has used data products from the Two Micron All
Sky Survey, which is a joint project of the University of
Massachusetts and the Infrared Processing and Analysis Center at the
California Institute of Technology, funded by NASA and NSF.

\vspace{5mm}
\facilities{FLWO 1.5m (TRES); MEarth}

\clearpage

\bibliographystyle{aasjournal}
\bibliography{masterref.bib}

\clearpage

\end{document}